\documentstyle[12pt]{article}
\newcommand{\be}[1]{\begin{equation} \label{(#1)}}
\newcommand{\ee}{\end{equation}}
\newcommand{\ba}[1]{\begin{eqnarray} \label{(#1)}}
\newcommand{\ea}{\end{eqnarray}}
\newcommand{\nn}{\nonumber}
\newcommand{\rf}[1]{(\ref{(#1)})}
\def \lsim {\mbox{${}^< \hspace*{-7pt} _\sim$}}
\def \gsim {\mbox{${}^> \hspace*{-7pt} _\sim$}}

\def\rp{$R_p \hspace{-1em}/\;\:$}
\def\rpm{R_p \hspace{-0.8em}/\;\:}
\def\rpt{$R_p \hspace{-0.85em}/\ \ $}
\def\et{$E_T \hspace{-1em}/\;\:$}
\def\pmb#1{\setbox0=\hbox{#1}%
  \kern-.015em\copy0\kern-\wd0
  \kern.03em\copy0\kern-\wd0
  \kern-.015em\raise.0233em\box0 }
\def \znbb {0\nu\beta\beta}
\def \tnbb {2\nu\beta\beta}
\def \emass {\langle m_{\nu} \rangle}
\def\bfr{\pmb{${r}$}}
\def\bfsgm{\pmb{${\sigma}$}}
\def\sir{({ \bfsgm_{i}^{~}} \cdot {\hat{\bfr}_{ij}^{~}} )}
\def\sjr{({ \bfsgm_{j}^{~}} \cdot {\hat{\bfr}_{ij}^{~}} )}
\def\si{{ \bfsgm_{i}^{~}}}
\def\sj{{ \bfsgm_{j}^{~}} }
\begin{document}

{\hfill \bf MPI-H-V 6-1995}
\bigskip
\begin{center}
{\bf  Supersymmetry and Neutrinoless Double Beta Decay}

\bigskip

{M. Hirsch\footnotemark[1], H.V. Klapdor-Kleingrothaus\footnotemark[2]
\bigskip

{\it
Max-Plank-Institut f\"{u}r Kernphysik, P.O. 10 39 80, D-69029,
Heidelberg, Germany}

\bigskip
S.G. Kovalenko\footnotemark[3]
\bigskip

{\it Joint Institute for Nuclear Research, Dubna, Russia}
}
\end{center}

\begin{abstract}
 Neutrinoless double beta decay ($\znbb$) induced by superparticle
exchange is investigated. Such a supersymmetric (SUSY) mechanism of
$\znbb$ decay arises within SUSY theories with R-parity
non-conservation (\rp).
We consider the minimal supersymmetric standard model (MSSM)
with explicit \rp terms in the superpotential (\rp MSSM).
The decay rate for the SUSY mechanism of $\znbb$ decay is calculated.
Numerical values for nuclear matrix elements for the
experimentally most interesting isotopes are calculated within pn-QRPA.
Constraints on the \rp MSSM parameter space are extracted from current
experimental half-life limits. The most stringent limits are
derived from data on $^{76}$Ge. It is shown that these constraints
are more stringent than those from other low-energy processes and
are competitive to or even more stringent than constraints expected
from accelerator searches.
\end{abstract}
\bigskip
\bigskip
\footnotetext[1]{MAHIRSCH@ENULL.MPI-HD.MPG.DE}
\footnotetext[2]{KLAPDOR@ENULL.MPI-HD.MPG.DE}
\footnotetext[3]{KOVALEN@NUSUN.JINR.DUBNA.SU}

\section{Introduction}
In the standard model (SM), since $B-L$ conservation is exact,
neutrinoless double beta decay ($\znbb$), which violates
lepton number by two units, is forbidden.
On the other hand  $B-L$ and $L$ violation is expected in theories
beyond the SM.
That is why $\znbb$ decay has long been recognized as a sensitive tool
to put theories beyond the SM to the test (for reviews see
\cite{gk90}, \cite{hax84}, \cite{doi85}).
A variety of mechanisms which may cause $\znbb$ have been studied
in the past.
The simplest and the most well-known possibility is via the exchange of
a Majorana neutrino between the decaying neutrons or due to
$(B-L)$-violating right-handed currents. $\znbb$ decay has not yet been
seen, but
limits on various model parameters can be deduced
(see ~\cite{hax84}, \cite{doi85} and references therein)
from its non-observation.

Recently impressive progress has been achieved in the experimental
investigation of double beta decay, both in the $2\nu\beta\beta$
and the $\znbb$ decay mode ~\cite{Klapdor}-~\cite{hdmo94}.
In the near future essential advance in this direction is expected.
Experimental lower bounds on $\znbb$-decay half-lives are often represented
in terms of an upper limit on the Majorana neutrino mass $\emass$.
At least one experiment currently in operation will reach a final
sensitivity of about $\emass$ $=$ ($0.1-0.2$) eV
and has already pushed the existing limit below $1$ eV \cite{hdmo94}.

   In view of the rising experimental sensitivity it is of great interest
to pursue a more comprehensive theoretical study of the possible
mechanisms of $\znbb$ decay.

In this work we investigate contributions to $\znbb$ decay
within supersymmetric (SUSY) theories with explicit R-parity breaking.
R-parity ($R_p$) is a discrete, multiplicative symmetry
defined as $R_p = (-1)^{3B+L+2S}$ ~\cite{fayet75}, where $S,\ B$ and $L$
are the spin, the baryon and the lepton quantum number.
The SM fields, including additional Higgs boson fields appearing
in the extended gauge models, have $R_p = +1$ while their superpartners
have $R_p = -1$. This symmetry has been imposed on
the minimal supersymmetric standard model (MSSM) (for a review
see \cite{Haber}) to ensure baryon number ($B$) and lepton number
($L$) conservation. However, neither gauge invariance nor
supersymmetry require $R_p$ conservation.
The question whether or not $R_p$ is a good symmetry of
the supersymmetric theory is a dynamical problem which might be
related to more fundamental physics at the Planck scale.
In general $R_p$ can be either broken explicitly
vacuum expectation value of the scalar superpartner of the
$R_p$-odd  isosinglet lepton field ~\cite{mv90}.

Supersymmetric models with $R_p$ non-conservation (\rp) have been
extensively discussed in the literature not only because of their great
theoretical interest, but also because they have interesting phenomenological
and cosmological implications.

Existing constraints on \rp SUSY theories are either direct from collider
experiments \cite{roy92} or indirect from low-energy processes \cite{bar89},
matter stability \cite{zwirn83}, ~\cite{weinberg82}, ~\cite{bm86}
and cosmology \cite{Cosmology}-\cite{dr94}.
We will discuss the former two constraints later on in more detail
in comparison with the bounds from $\znbb$ decay.

Recently \rp SUSY models have been analysed in connection
with current and forthcoming collider experiments.
The \rp SUSY gives rise to spectacular signatures
of events in collider detectors which would yield a very clean signal
for supersymmetry. Consequently,
expected sensitivities of experiments at HERA \cite{but93}, the TEVATRON
\cite{bae94}, LEP 200 ~\cite{grt93} and the LHC ~\cite{dgr94} have been
analysed recently.
In the present paper we pay special attention to comparing the capability
of $\znbb$ decay experiments with the collider experiments in
establishing better constraints on \rp SUSY models.

The rest of the paper is organized as follows. In the next section, we
specify the minimal supersymmetric standard model with $R_p$ non-conservation.
Section 3 discusses the basic diagrams inducing $\znbb$ decay
in the \rp MSSM.
The effective low-energy Lagrangian as well as the different
lepton number violating parameters are defined. Section 4 outlines the
procedure for obtaining nucleon matrix elements from the quark currents
in the non - relativistic impulse approximation. Section 5 then deals
with the numerical calculation of the relevant nuclear structure matrix
elements. We briefly summarize the main features of the nuclear structure
model, before presenting numerical results for those isotopes which are
currently the experimentally most promising. Special attention is paid
to a discussion of the theoretical uncertainties of the nuclear
structure calculation. In section 6,  on the basis of the
current experimental limit on the half-life of $^{76}$Ge \cite{hdmo94},
we analyse constraints on the supersymmetric parameter space imposed
by the non-observation of $\znbb$ decay. We have found that these limits
are more stringent than those from other low-energy processes and also
more stringent than those expected from experiments
with the ZEUS detector at HERA \cite{but93}.
We then close with a short summary and outlook.

\section{Minimal Supersymmetric Standard Model with $R$-parity
						Non-conservation}
In the following we will use the MSSM extended by inclusion of
the explicit $R$-parity non-conserving terms (\rp) into the superpotential.
This model has the MSSM field content and is completely specified
by the standard $SU(3)\times SU(2)\times U(1)$ gauge couplings
as well as by the low-energy superpotential and "soft" SUSY
breaking terms \cite{Haber}.
The most general gauge invariant form of the superpotential is
\be{sup_gen}
           W = W_{R_p} + W_{\rpm}.
\ee
The $R_p$ conserving part has the standard MSSM form
\be{R_p-cons} 
         W_{R_p} = h_L H_1 L {\bar E} + h_D H_1 Q {\bar D}
                   - h_U H_2 Q {\bar U} - \mu H_1 H_2.
\ee
We use notations $L$,  $Q$ for lepton and quark
doublet superfields and ${\bar E}, \ {\bar U},\  {\bar D}$ for lepton and
{\em up}, {\em down} quark singlet  superfields;
$H_1$ and $H_2$ are the Higgs doublet superfields
with a weak hypercharge $Y=-1, \ +1$, respectively.
Summation over generations is implied.
For simplicity generation indices of fields and Yukawa coupling
constants  $h_L, \ h_U, \ h_D$  are suppressed.
The mass-mixing parameter $\mu$ is a free parameter describing
mixing between the Higgs bosons $H_{1}$-$H_{2}$
as well as between higgsinos $\tilde H_1$-$\tilde H_2$.

The $R_p$ violating part of the superpotential \rf{sup_gen}
can be written as ~\cite{dh87}, ~\cite{hs84},
\be{R-viol} 
W_{\rpm} = \lambda_{ijk}L_i L_j {\bar E}_k + \lambda'_{ijk}L_i Q_j {\bar D}_k
+ \lambda''_{ijk}{\bar U}_i {\bar D}_j {\bar D}_k ,
\ee
where indices $i,j, k$ denote generations, and the fields have been defined
so that the bilinear lepton number violating operators  $L_i H_2$ ~\cite{hs84}
have been rotated away.
The coupling constants $\lambda$ ($\lambda''$) are antisymmetric in the first
(last) two indices. The first two terms lead to lepton number violation,
while the last one violates baryon number conservation.

Proton stability forbids the simultaneous presence of lepton and baryon number
violating terms in the superpotential \cite{zwirn83} (unless the couplings
are very small).  Therefore, only $\lambda$, $\lambda'$ {\it or} $\lambda''$
type interactions can be present.
There may exist an underlying
discrete symmetry in the theory which allows either the first or
the second set of couplings ~\cite{weinberg82},
An example of such a symmetry, which forbids
baryon number violating couplings but allows lepton violating ones,
is  given by the transformation rules ~\cite{dr91}
\be{symmetry}
\left(Q, \bar U,  \bar D  \right)
\longrightarrow - \left(Q, \bar U,  \bar D  \right),
\left(L, \bar E,  H_{1,2} \right)
\longrightarrow  + \left(L, \bar E,  H_{1,2} \right) .
\ee
This discrete symmetry can be justified on a more fundamental level
of Planck scale physics.
It has been shown to be compatible with the ordinary
$SU(5)$ ~\cite{hs84} and "flipped" $SU(5)\times U(1)$ ~\cite{bh89}
grand unification (GUT) scenarios as well as with phenomenologically viable
superstring theories ~\cite{sstr}.

Neutrinoless double beta decay, which is the main subject of the present paper,
requires lepton number violating interactions.
Therefore  we bind ourselves to the \rp MSSM with
lepton number violation ($\lambda\neq 0,\   \lambda'\neq 0$)
and baryon number conservation ($\lambda'' = 0$). The Lagrangian of
this model possesses the discrete symmetry eq. \rf{symmetry}.
 Apparently, $\znbb$ can probe only the first generation
lepton number violating coupling $\lambda'_{111}$ because only
the first generation fermions $u, d, e$ are involved in this process.

In addition to proton decay constraints on \rp couplings
there are also constraints which follow from cosmological arguments,
requiring that the baryon asymmetry generated at the GUT scale is not
washed out by $B-L$ violating interactions present in eq. \rf{R-viol}.
These cosmological constraints have been thought to affect all \rp couplings
$\lambda, \lambda', \lambda''\ll 10^{-7}$, making
these models phenomenologically not interesting.
These arguments, however, were proved to be strongly
model dependent ~\cite{mv87},
bounds can be evaded in perfectly reasonable scenarios of matter
genesis \cite{dr94}.

The effect of "soft" supersymmetry breaking can be parametrized
at the Fermi scale as a part of the scalar potential:
\ba{V_soft} 
V_{soft}= \sum_{i=scalars}^{}  m_{i}^{2} |\phi_i|^2 +
h_L A_L H_1 \tilde L \tilde{\bar E} + h_D A_D H_1 \tilde Q \tilde{\bar D}
- h_U A_U H_2 \tilde Q \tilde{\bar U}  -&& \\
\nn
- \mu B H_1 H _2 + \mbox{ h.c.}&&
\ea
and a "soft"  gaugino mass term
\be{M_soft} 
{\cal L}_{GM}\  = \ - \frac{1}{2}\left[M_{1}^{} \tilde B \tilde B +
 M_{2}^{} \tilde W^k \tilde W^k  + M_{3}^{} \tilde g^a \tilde g^a\right]
 -   \mbox{ h.c.}
\ee
As usual, $M_{3,2,1}$ are the masses of the $SU(3)\times
SU(2)\times U(1)$ gauginos $\tilde g, \tilde W, \tilde B$ and $m_i$
are the masses of scalar fields. $A_L,\ A_D, \ A_U$ and $B$ are
trilinear and bilinear "soft" supersymmetry breaking parameters.
All these quantities are free SUSY model parameters which due to
the renormalization effect depend on the energy scale $\Lambda$.

Considering a GUT scenario within the MSSM one can
claim the following unification conditions at the GUT scale $ \Lambda\sim M_X$:
\be{boundary1}  
 A_U(M_X) = A_D(M_X) = A_L(M_X) = A_{0},
\ee
\be{boundary2}  
 m_{L}(M_X) = m_{E}(M_X) = m_{Q}(M_X) = m_{U}(M_X) = m_{D}(M_X) = m_{0},
\ee
\be{boundary3}  
 M_{1}(M_X) = M_{2}(M_X) = M_{3}(M_X) = m_{1/2},
\ee
\be{boundary4} 
g_{1}^{}(M_X) = g_{2}^{}(M_X) = g_{3}^{}(M_X) =  g_{GUT}^{},
\ee
where $g_{3}^{}, g_{2}^{}, g_{1}^{}$ are the
$SU(3)\times SU(2)\times U(1)$ gauge coupling constants
equal to $g_{GUT}^{}$  at the unification scale $M_X$.

   At the Fermi scale $\Lambda\sim M_W$ these parameters
can be evaluated on the basis of the MSSM renormalization group equations
(RGE) \cite{RGE},\cite{SfermionMass}.
We assume that the \rp Yukawa coupling
constants $\lambda, \lambda', \lambda''$ are small enough
to be neglected in these equations.
Equation \rf{boundary3} implies at $\Lambda\sim M_W$
\be{M1_M2}
M_1 = \frac{5}{3} \tan^2\theta_W \cdot M_2, \ \ \ M_2 \simeq 0.3 m_{\tilde g}.
\ee
Here $m_{\tilde g} = M_3$ is the gluino mass.

Now the model is completely specified and we can deduce
the interaction terms of the  \rp  MSSM - Lagrangian  relevant for
neutrinoless double beta decay.

Write down these  interaction terms explicitly. Note that in the
following  we use  for fermion fields the 4-component Dirac
bispinor notation.

The lepton number violating part of the Lagrangian can be obtained directly
from the superpotential \rf{R-viol}. It  has the form
        \ba{Lqqe}
        {\cal L}_{\rpm} = &-& \lambda'_{111}\left[
	(\bar{u}_L \ \bar{d}_R)\cdot
	\mbox{$ \left( \begin{array}{cc}
	e_{R}^{c}\\
	-\nu_{R}^{c}
 	\end{array} \right) $}\ \tilde{d}_R
	+
	(\bar{e}_L\ \bar{\nu}_L)\ d_R\cdot
	\mbox{$ \left( \begin{array}{cc}
	\tilde{u}_{L}^{\ast}\\
	-\tilde{d}_{L}^{\ast}
 	\end{array} \right) $} + \right. \\ \nn
	&+& \left. (\bar{u}_L\ \bar{d}_L)\ d_R \cdot
	\mbox{$ \left( \begin{array}{cc}
	\tilde{e}_{L}^{\ast}\\
	-\tilde{\nu}_{L}^{\ast}
 	\end{array} \right) $}
	 + h.c. \right]
	\ea
The  Lagrangian terms corresponding to  gluino ${\cal L}_{\tilde{g}}$
and neutralino ${\cal L}_{\chi}$ interactions with fermions
$\psi = \{u, d, e\}$, $q = \{u, d\}$ and their superpartners
$\tilde \psi = \{\tilde u, \tilde d, \tilde e\}$,
$\tilde q =\{\tilde u, \tilde d\}$
are \cite{Haber}
 \ba{gluino}
                {\cal L}_{\tilde g} = - \sqrt{2} g_3
        \frac{{\bf \lambda}^{(a)}_{\alpha \beta}}{2}
        \left( \bar q_L^{\alpha} \tilde g \tilde q_L^{\beta}
                          - \bar q_R^{\alpha} \tilde g \tilde q_R^{\beta}
                        \right)
			+ h.c.,
\ea
\ba{neutralino}
 	{\cal L}_{\chi} = \sqrt{2} g_2 \sum_{i=1}^4 \left(
                        \epsilon_{L i}(\psi) \bar \psi_L
				\chi_i \tilde \psi_L
        +  \epsilon_{R i}(\psi) \bar \psi_R \chi_i \tilde \psi_R\right)
			+ h.c.
\ea
Here
${\bf \lambda}^{(a)}$ are $3\times 3$ Gell-Mann matrices ($a = 1,..., 8$).
Neutralino coupling constants are defined as \cite{Haber}
\ba{eps}
        \epsilon_{L i}(\psi) &=& - T_3(\psi) {\cal N}_{i2} +
                                \tan \theta_W \left(T_3(\psi)
                        -  Q(\psi)\right) {\cal N}_{i1},\\
        \epsilon_{R i}(\psi) &=& Q(\psi) \tan \theta_W {\cal N}_{i1}.
\ea

Here $Q(\psi) $ and $ \ T_3(\psi)$ are the electric charge and weak
isospin of the field $\psi$.

Coefficients $N_{ij}$ are elements of
the orthogonal neutralino mixing matrix which diagonalizes
the neutralino mass matrix.
In the \rp MSSM the neutralino mass matrix is identical to
the MSSM one \cite{Haber} and in the basis of fields
($\tilde{B}, \tilde{W}^{3}, \tilde{H}_{1}^{0},  \tilde{H}_{2}^{0}$)
has the form:
\be{MassM}  
                M_{\chi} =  \left(
                        \begin{array}{cccc}
 M_1 & 0 & -M_Z s^{}_W c_\beta & M_Z s^{}_W s_\beta \\
 0   & M_2 & M_Z c^{}_W c_\beta & -M_Z c^{}_W s_\beta\\
 -M_Z s^{}_W c_\beta &  M_Z c^{}_W c_\beta & 0 & -\mu  \\
  M_Z s^{}_W s_\beta & -M_Z c^{}_W s_\beta & -\mu & 0 \\
 \end{array}
                     \right),
\ee
where $c^{}_W = \cos\theta_W$, $s^{}_W = \sin\theta_W$,
$t^{}_W = \tan \theta_W$, $s_\beta = \sin\beta$,
$c_\beta = \cos\beta$.
The angle $\beta$ is defined as $\tan\beta = <H_{2}^{0}>/<H_{1}^{0}>$.
Here $ <H_{2}^{0}>$ and $<H_{1}^{0}>$ are vacuum expectation values
of the neutral components $H_{2}^{0}$ and $H_{1}^{0}$ of the Higgs
doublet fields with weak hypercharges
$Y(H_{2}^0)  = +1$ and $Y(H_{1}^{0}) = -1$, respectively.
The mass parameters $M_1, M_2$ are related to the
gluino mass $m_{\tilde g}$  according to eq. \rf{M1_M2}.

By diagonalizing the mass matrix \rf{MassM} one can obtain
four neutralinos $\chi_i$ with masses $m_{\chi_i}$ and
the field content
\be{admix}
\chi_i = {\cal N}_{i1} \tilde{B} +  {\cal N}_{i2}  \tilde{W}^{3} +
{\cal N}_{i3} \tilde{H}_{1}^{0} + {\cal N}_{i4} \tilde{H}_{2}^{0}.
\ee
Recall again that we use notations
$\tilde{W}^{3}$, $\tilde{B}$ for neutral $SU(2)_L \times U(1)$
gauginos and  $\tilde{H}_{2}^{0}$, $\tilde{H}_{1}^{0}$
for higgsinos which are the superpartners of the two neutral Higgs boson
fields $H_1^0$ and $H_2^0$.

 We apply a diagonalization by means of a real orthogonal matrix
${\cal N}$.
Therefore the coefficients ${\cal N}_{ij}$ are real and
masses $m_{\chi_i}$ are either positive or negative.
The sign of the mass coincides with the CP-parity of the corresponding
neutralino mass eigenstate $\chi_i$.
If necessary, a negative mass can be always made positive by
a redefinition  \cite{Gunion} of  the neutralino field $\chi_i$.
It leads to a redefinition of the relevant mixing coefficients
${\cal N}_{ij} \rightarrow i\cdot {\cal N}_{ij}$.

The lightest neutralino is commonly assumed to be
the lightest supersymmetric particle (LSP). That is true
in almost all phenomenologically viable SUSY models with $R_p$ conservation.
For \rp SUSY models this is a very non-trivial assumption,
since the cosmological constraints ~\cite{EHNOS} requiring
the LSP to be colour and electrically neutral, no longer apply ~\cite{camp91}.
A priori, the LSP could be any superparticle
in \rp SUSY models.
However, the RGE analysis in minimal supergravity models suggests
that the LSP is a neutralino if \rp couplings are reasonably small
the lightest neutralino. If $R_p$ is conserved the LSP is
a stable particle. Otherwise it decays into ordinary matter.

Squarks ${\tilde u}_{L,R},\ {\tilde d}_{L,R}$ and
selectron $\tilde e_{L,R}$ in eqs. \rf{gluino}, \rf{neutralino}
are with a good precision mass eigenstates.
Possible $\tilde f_L-\tilde f_R$-mixing for
the first generation of squarks and sleptons are negligible
due to the smallness of the relevant Yukawa couplings.
In this case the MSSM mass formulas
can be written as \cite{Haber}, ~\cite{RGE} \cite{SfermionMass}
\ba{SfermionMass}
 m^2_{\tilde{e}_L} &=& m^2_0 + 0.07 m_{\tilde g}^2
        + \frac{1}{2}\cos\!2\beta M_Z^2 (2 \sin^2\theta_W - 1), \\
 m^2_{\tilde{e}_R} &=&  m^2_0 + 0.02 m_{\tilde g}^2
       -             \cos\!2\beta M_Z^2 \sin^2\theta_W,\\
m^2_{\tilde{u}_L} &=&  m^2_0 + 0.83  m_{\tilde g}^2
      + \frac{1}{2}\cos\!2\beta M_Z^2 (1- \frac{4}{3} \sin^2\theta_W ),\\
m^2_{\tilde{d}_L} &=& m^2_0 + 0.83 m_{\tilde g}^2
      - \frac{1}{2}\cos\!2\beta M_Z^2 (1 - \frac{2}{3} \sin^2\theta_W ), \\
m^2_{\tilde{u}_R} &=&  m^2_0 + 0.77 m_{\tilde g}^2
                       + \cos\!2\beta  M_Z^2  \frac{2}{3} \sin^2\theta_W, \\
\label{(smass_end)}
m^2_{\tilde{d}_R} &=&  m^2_0 +  0.76 m_{\tilde g}^2
                       - \cos 2\beta  M_Z^2  \frac{1}{3} \sin^2\theta_W.
\ea
Here,  $m_{\tilde g}$ is the gluino mass and $m_0$ is the common sfermion
mass at the unification scale (see \rf{boundary2}).
{}From the eqs. \rf{SfermionMass}-\rf{smass_end} one can estimate in the region
$m_{(\tilde u)_L}^2, m_{(\tilde d)_R}^2\lsim 300$GeV
that $m_{(\tilde u)_L}^2 \approx m_{(\tilde d)_R}^2$.
This approximate relation will be helpful for understanding
the results of our numerical analysis in Sect. 6.

Having specified the Lagrangian interaction terms \rf{Lqqe} - \rf{neutralino}
and the mass eigenstates $\tilde g, \chi_i, \tilde q, \tilde e$
involved in the interactions we can construct diagrams describing the \rp MSSM
contribution to the neutrinoless double beta decay.

In principle, in the \rp MSSM a small neutrino mass of Majorana type
may arise radiatively ~\cite{hs84},~\cite{Rad_Neutmass}.  On the other hand
in some GUT scenarios based on large gauge groups like $SO(10)$ there
might be also heavy Majorana fermions having non-negligible $SU(2)_L$
components. It is often identified with the heavy Majorana neutrino $N$.
In the presence of either the light Majorana neutrinos  or the heavy ones
(or both) $\znbb$ decay can be induced by the conventional
neutrino mass mechanism. In the present paper we concentrate on the \rp SUSY
mechanism of $\znbb$ decay.
The effect of additional contributions from the neutrino mass mechanism
is investigated in the case of heavy Majorana neutrino exchange.
(Inclusion of the light Majorana neutrino contribution will not change
the results of our analysis.)

\section{Low-energy $\Delta L_e = 2$ -  Effective Lagrangian }

The basic diagram corresponding to the neutrinoless double beta decay
at the nucleon
level is presented in fig.1(a). Two neutrons from the initial nucleus $A_i$
after interaction transform into two protons of the final nucleus $A_f$
emitting two electrons. Apparently this process breaks the electron
lepton number by two units, $\Delta L_e = 2$.  At the quark level it can be
induced  by the subprocess with two initial d-quarks and two final u-quarks
accompanied by two electrons as shown in Fig 1(b).

The conventional mass mechanism with Majorana neutrino
exchange is presented in fig. 2(a).
In the following we consider the \rp MSSM contribution to $\znbb$ decay.
Starting from the fundamental interactions \rf{Lqqe}-\rf{neutralino}
we have found
\cite{HKK} the complete set of diagrams presented in fig. 2(b, c)
which contribute to this subprocess.

The supersymmetric mechanism of $\znbb$ decay was first proposed
by Mohapatra \cite{Mohapatra}
and later studied in more details by Vergados \cite{Vergados1}.
In these papers \cite{Mohapatra}, \cite{Vergados1} only three diagrams
similar to those in fig.2(b) were considered.
Instead of neutralinos $\chi_i$, which are actual mass eigenstates
in the MSSM, the consideration of refs. \cite{Mohapatra},~\cite{Vergados1}
used Z-ino ($\tilde Z$) and photino ($\tilde \gamma$) fields
in intermediate states. $\tilde Z$ and $\tilde \gamma$
can be mass eigenstates only at special values of parameters of
the neutralino mass matrix.
In general, these fields are not mass eigenstates.
Furthermore, it was recently realized that a cosmologically viable
lightest neutralino is very likely B-ino dominant ~\cite{dmlimit}.
Using in the intermediate states such fields which
are not mass eigenstates leads to neglecting diagrams with
mixed intermediate states when, for instance,
$\tilde Z$ turns to $\tilde \gamma$ due to the mixing
proportional to the relevant entry of the neutralino mass matrix.
The effect of mixing is taken into account completely in the set of diagrams
displayed in fig.2(b,c) with all neutralino mass eigenstates $\chi_i$ involved.

In the case of $0\nu\beta\beta$ decay when momenta of external particles
are much  smaller
than
intermediate particle masses one can treat
interactions in fig. 1(b) and fig. 2 as point-like.  A suitable formalism in
this case is the effective Lagrangian approach.

Define the effective Lagrangian ${\cal L}^{\Delta L_e =2}_{eff}(x)$ as

\ba{DefL}
<f|S - 1|i>\ &=&\ i \int d^4 x <f| {\cal L}^{\Delta L_e =2}_{eff}(x)|i> + \\
\nn
      &+& \mbox{high  orders of  perturbation theory.}
\ea

Thus, ${\cal L}^{\Delta L_e =2}_{eff}(x)$ corresponds to the lowest
order operator structure having non-vanishing matrix elements of the form

\ba{LLL}
<u u e e|{\cal L}^{\Delta L_e =2}_{eff}(x)|d d> \neq 0 .
\ea

It is now straightforward to find the operators in
the effective Lagrangian which correspond to the diagrams in fig. 2(b,c).
The result is \cite{HKK}
\ba{Leff}
&&{\cal L}^{\Delta L_e =2}_{eff}(x)\ =\
8 \pi \alpha_2 \lambda^{'2}_{111} \sum_{i=1}^{4} \frac{1}{m_{\chi_i}}\ \left[
\frac{\epsilon_{L i}^2(e)}{m_{\tilde e_L}^4}
(\bar u_L^{\alpha} d_{R \alpha})(\bar u_L^{\beta} d_{R \beta})
(\bar e_L e^{\bf c}_{R}) + \right. \\ \nn
&&+ \left.\frac{\epsilon_{L i}^2(u)}{m_{\tilde u_L}^4}
(\bar u_L^{\alpha} u^{\bf c}_{R \beta})(\bar e_L d_{R \alpha})
(\bar e_L d_{R}^{\beta}) +
\frac{\epsilon_{R i}^2(d)}{m_{\tilde d_R}^4}
(\bar u_L^{\alpha} e^{\bf c}_R)(\bar u_L^{\beta} e^{\bf c}_R)
(\overline{ d^{\bf c}}_{L \alpha} d_{R \beta}) + \right.\\ \nn
&+& \left. \left(
\frac{\epsilon_{L i}(u)\epsilon_{R i}(d)}{m_{\tilde u_L}^2 m_{\tilde d_R}^2}
+ \frac{\epsilon_{L i}(u)\epsilon_{L i}(e)}{m_{\tilde u_L}^2 m_{\tilde e_L}^2}
+\frac{\epsilon_{L i}(e)\epsilon_{R i}(d)}{m_{\tilde e_L}^2 m_{\tilde d_R}^2}
\right) (\bar u_L^{\alpha} d_R^{\beta})(\bar u_{L \beta} e^{\bf c}_R)
(\overline{e}_L d_{R \alpha})
\right]\ + \\ \nn
&&+\  \lambda^{'2}_{111} \frac{8 \pi \alpha_s}{{m_{\tilde g}}}
\frac{{\bf \lambda}^{(a)}_{\alpha \beta}}{2}
\frac{{\bf \lambda}^{(a)}_{\gamma \delta}}{2}
 \left[
\frac{1}{m_{\tilde u_L}^4}
(\bar u_L^{\alpha} u^{{\bf c} \gamma} _{R})(\bar e_L d_R^{\beta})
(\bar e_L d_R^{\delta}) + \right.\\ \nn
&+& \left.\frac{1}{m_{\tilde d_R}^4}
(\bar u_L^{\alpha} e^{\bf c}_R)(\bar u_L^{\gamma} e^{\bf c}_R)
(\overline{ d^{\bf c}}_L^{\beta} d^{\bf c\ \delta}_{R})
- \frac{1}{m_{\tilde d_R}^2 m_{\tilde u_L}^2}
(\bar u_L^{\alpha} d_R^{\delta})(\bar u_L^{\gamma} e^{\bf c}_R)
(\overline{e}_L d_{R}^{\beta})
\right]\  \
\ea
Gauge coupling constants
$\alpha_2 = g_{2}^{2}/(4\pi)$ and $\alpha_s = g_{3}^{2}/(4\pi)$
are running coupling constants which should be estimated at
the proper energy scale $\Lambda$.
One can see from the diagrams in fig.2 that the typical scale at which the
$\tilde g - \tilde q - q$ and $\chi - \tilde q - q$ interactions
occur is of the order of the gluino or neutralino mass.
In the mass region which will be analysed numerically in the subsequent
sections we may take approximately these couplings
at the Z-boson pole and use their values from ref.  \cite{CouplingConst}:
\ba{Coup}
\alpha_s(M_Z) = 0.127, \ \ \ \alpha_2(M_Z) = 0.0337.
\ea

The Lagrangian \rf{Leff} has terms in a form which do not
allow a direct application of the non-relativistic impulse approximation
for further calculation
of the $\znbb$ reaction matrix element.
One should rearrange the right hand side of eq. \rf{Leff}  in
the form of a product
of two colour-singlet quark currents and the leptonic current.
It can be accomplished by a Fierz rearrangement procedure and subsequent
extraction of color-singlets from the product of two colour-triplet
and colour-antitriplet  quark fields.
The final result is
\ba{Leta}
{\cal L}^{\Delta L_e =2}_{eff}(x)\ &=&\
\frac{G_F^2}{2}\cdot m_P^{-1}\left[(\eta_{\tilde g} + \eta_{\chi})
(J_{PS}J_{PS} - \frac{1}{4} J_T^{\mu\nu} J_{T \mu\nu}) \ + \right. \\ \nn
&+& \left. (\eta_{\chi \tilde e} + \eta'_{\tilde g} - \eta_{\chi \tilde f})
J_{PS} J_{PS} + \eta_N J_{VA}^{\mu}J_{VA \mu}\right]
(\bar e (1 + \gamma_5) e^{\bf c}),
\ea
The last term corresponds to the heavy Majorana neutrino contribution
described by the diagram in fig. 2(a) which we also consider in
the present paper as discussed at the end of section 2.

The lepton number violating parameters are defined as follows
\ba{eta}
\eta_{\tilde g} &=& \frac{2 \pi \alpha_s}{9}
\frac{\lambda^{'2}_{111}}{G_F^2 m_{\tilde d_R}^4} \frac{m_P}{m_{\tilde
g}}\left[
1 + \left(\frac{m_{\tilde d_R}}{m_{\tilde u_L}}\right)^4\right]\\
\eta_{\chi} &=& \frac{ \pi \alpha_2}{6}
\frac{\lambda^{'2}_{111}}{G_F^2 m_{\tilde d_R}^4}
\sum_{i=1}^{4}\frac{m_P}{m_{\chi_i}}
\left[
\epsilon_{R i}^2(d) + \epsilon_{L i}^2(u)
\left(\frac{m_{\tilde d_R}}{m_{\tilde u_L}}\right)^4\right]\\
\eta_{\chi \tilde e} &=& 2 \pi \alpha_2
\frac{\lambda^{'2}_{111}}{G_F^2 m_{\tilde d_R}^4}
\left(\frac{m_{\tilde d_R}}{m_{\tilde e_L}}\right)^4
\sum_{i=1}^{4}\epsilon_{L i}^2(e)\frac{m_P}{m_{\chi_i}},\\
\eta'_{\tilde g} &=& \frac{4 \pi \alpha_s}{9}
\frac{\lambda^{'2}_{111}}{G_F^2 m_{\tilde d_R}^4}
\frac{m_P}{m_{\tilde g}} \left(\frac{m_{\tilde d_R}}{m_{\tilde
u_L}}\right)^2,\\
\label{(eta_end)}
\eta_{\chi \tilde f} &=& \frac{\pi \alpha_2 }{3}
\frac{\lambda^{'2}_{111}}{G_F^2 m_{\tilde d_R}^4}
\left(\frac{m_{\tilde d_R}}{m_{\tilde e_L}}\right)^2
\sum_{i=1}^{4}\frac{m_P}{m_{\chi_i}}
\left[\epsilon_{R i}(d) \epsilon_{L i}(e)  + \right.\\ \nn
&+& \left.\epsilon_{L i}(u) \epsilon_{R i}(d)
\left(\frac{m_{\tilde e_L}}{m_{\tilde u_L}}\right)^2
+ \epsilon_{L i}(u) \epsilon_{L i}(e)
\left(\frac{m_{\tilde d_R}}{m_{\tilde u_L}}\right)^2
\right],\\
\label{(m_N)}
\eta_N &=& \frac{m_P}{<m_N>},
\ea
where $<m_N>$ is the {\em effective} heavy Majorana neutrino mass
(for definition see ~\cite{doi85}).

Colour-singlet hadronic currents have the form
\ba{Hcurr}
J_{PS} &=& \bar u^{\alpha} (1 + \gamma_5) d_{\alpha},\\
J_T^{\mu \nu} &=& \bar u^{\alpha} \sigma^{\mu \nu}(1 + \gamma_5) d_{\alpha},\\
J_{AV}^{\mu} &=& \bar u^{\alpha} \gamma^{\mu}(1 - \gamma_5) d_{\alpha},
\label{(JT)}
\ea
where $\sigma^{\mu \nu} = \frac{i}{2} [\gamma^{\mu}, \gamma^{\nu}]$.

Now we can consider nucleon matrix elements of these quark currents
which provide us with certain information about quark states
in the nucleon.
Further we will use in Sect. 5 the proper nuclear wave functions
to describe nucleon states in the nucleus. Applying this standard
two-step-procedure we will obtain the reaction matrix element
for the $\znbb$ decay.

\section{From quark to nuclear level}

Adopting the above mentioned two-step-procedure let us write down the
$\znbb$ decay matrix element ${\cal R}_{\znbb}$ corresponding to
the effective Lagrangian eq. \rf{Leta}. Using the
general formula eq. \rf{matr}
one can write
\ba{R_0nu}
{\cal R}_{\znbb} &=& \frac{G_{F}^{2}}{\sqrt{2}} m_{P}^{-1} C_{0\nu}^{-1}
\left\{\bar{e}(1 + \gamma_5) e^{\bf c} \right\}\times \\ \nn
&&\left[
\left(\eta_{\tilde g} + \eta_{\chi} \right)\Big<F| \Omega_{\tilde q} |I \Big> +
\left(\eta_{\chi \tilde e} + \eta'_{\tilde g} - \eta_{\chi \tilde f}
\right)\Big<F| \Omega_{\tilde f} |I \Big>
+ \eta_N \Big<F| \Omega_N |I \Big> \right].
\ea
The heavy Majorana neutrino exchange contribution has been included
in the last term of this equation. It corresponds to the last term
of eq. \rf{Leta}.

We have introduced  transition operators  $\Omega_i$
(see eq. \rf{O_def} in Appendix) which are useful for separating
the particle physics part of the calculation from
the nuclear physics one. The transition operators contain information
about the underlying interactions at the quark level \rf{Leta}
and quark states inside the nucleon. They  are independent of
the initial $|I>$ and the final $<F|$ nuclear states.
To calculate the nuclear matrix elements
in eq. \rf{R_0nu} one should use nuclear model wave functions.

Nuclear matrix elements of the transition operators
$\Omega_{\tilde q}, \Omega_{\tilde f}$ and $\Omega_N$ describe
transitions induced by quark colour singlet currents \rf{Hcurr}-\rf{JT}
in the first, second and third terms of eq. \rf{Leta}.
Diagrams in fig. 2 with the intermediate states $\{W - N - W\}$,
$\{\tilde u(\tilde d) - \chi,\tilde g - \tilde u(\tilde d)\}$ and
$\{\tilde u(\tilde d) - \chi,\tilde g - \tilde d(\tilde u);
\tilde e - \chi - \tilde e,\tilde q\}$
contribute to operators $\Omega_N$, $\Omega_{\tilde q}$ and
$\Omega_{\tilde f}$, respectively.

The relevant formula for calculating the transition operators
in the non-relativistic impulse approximation (NRIA) \rf{NRIA} require
the nucleon matrix elements of these currents.

Now we turn to the derivation of nucleon matrix elements of
the colour singlet quark  currents \rf{Hcurr}-\rf{JT} using results
of ref. \cite{Adler}.
The relevant matrix elements are,
\ba{PS}
<P(p)| \bar u d|N(p')> &=& F_S^{(3)}(q^2)\cdot \bar N(p) \tau_{+} N(p'),\\
<P(p)| \bar u \gamma_5 d|N(p')> &=& F_P^{(3)}(q^2)\cdot \bar N(p)
				    \gamma_5\tau_{+} N(p'),\\
<P(p)| \bar u \sigma^{\mu \nu}(1 + \gamma_5) d|N(p')>
	&=& \bar N(p) \left(J^{\mu\nu} +
\frac{i}{2} \epsilon^{\mu\nu\rho\sigma} J_{\rho\sigma}\right) \tau_{+} N(p'),\\
<P(p)|\bar u \gamma^{\mu}(1 - \gamma_5) d |N(p')>
&=& \bar N(p)\gamma^{\mu}\left(F_V(q^2) - F_A(q^2)\gamma_5\right)\tau_{+}
N(p'),
\ea
where $q = p - p'$ and the tensor structure is defined as

\ba{JJJ}
J^{\mu\nu} = T_1^{(3)}(q^2) \sigma^{\mu\nu} + \frac{i T_2^{(3)}}{m_P}
\left(\gamma^{\mu} q^{\nu} -  \gamma^{\nu} q^{\mu}\right) +
\frac{T_3^{(3)}}{m_P^2}
\left(\sigma^{\mu\rho} q_{\rho} q^{\nu} -  \sigma^{\nu\rho} q_{\rho}
q^{\mu}\right).
\ea

For all form factors $F_{V,A}(q^2), F_{S,P}^{(3)}(q^2),
T_i^{(3)}(q^2)$ we take following ref. \cite{Vergados2} a dipole form
\ba{dip}
\frac{F_{V,A}(q^2)}{f_{V,A}} =
\frac{F_{S,P}^{(3)}(q^2)}{F_{S,P}^{(3)}(0)} =
\frac{T_i^{(3)}(q^2)}{T_i^{(3)}(0)} = \left(1 - \frac{q^2}{m_A^2} \right)^{-2}
\ea

with $m_A = 0.85$GeV and $f_V\approx 1,\ f_A\approx 1.261$.
Form factor normalization values $F_{S,P}^{(3)}(0)$, $T_i^{(3)}(0)$
were calculated in ref. \cite{Adler} and are given in Table 1.

\begin{table}[t]
{Table 1: Nucleon form factor normalizations (at $q^2 = 0$) as calculated in
\cite{Adler}.}\\[5mm]
\begin{tabular}{|c|c|c|c|c|c|}\hline
Set & $F_S^{(3)}$ & $F_P^{(3)}$ & $T_1^{(3)}$ & $T_2^{(3)}$ & $T_3^{(3)}$\\
\hline
A) Bag Model  & 0.48 & 4.41 & 1.38 & -3.30 & -0.62\\
&&&&&\\
\hline
B) Non-relativistic  & 0.62 & 4.65 & 1.45 & -1.48 & -0.66\\
quark model&&&&&\\
\hline
\end{tabular}
\end{table}

Using formulas \rf{PS}-\rf{JJJ} we may derive the non-relativistic limit
$m_P \gg |\vec p|$ for
nucleon matrix elements of the three combinations of the quark currents
in \rf{Leta}. Keeping all terms up to order $q^2$ in
the non-relativistic expansion we find from eq. \rf{NRIA}
the relevant transition operators
$\Omega_{\tilde q}, \Omega_{\tilde f}, \Omega_N$ for
two outgoing electrons in the S-wave state,
\ba{Omega}
\Omega_{\tilde q} &=&
{\frac{m_P}{ m_e}}\Big\{ \alpha_V^{(0)} \Omega_{F,N} +
\alpha_A^{(0)}\Omega_{GT,N} + \alpha_V^{(1)} \Omega_{F'}  +
\alpha_A^{(1)} \Omega_{GT'} + \alpha_T \Omega_{T'} \Big\},\\
\label{(Omega_f)}
\Omega_{\tilde f} &=& \Omega_{\tilde q}(T_i = 0),\\
\label{(Omega_N)}
\Omega_N &=&
{\frac{m_P}{ m_e}}\Big\{ \left(\frac{f_V}{f_A}\right)^2 \Omega_{F,N} -
                        \Omega_{GT,N} \Big\},
\ea
where partial transition operators are
\ba{O_i}
\Omega_{GT,N} &=&  \sum_{i \ne j} \tau_{+}^{(i)} \tau_{+}^{(j)}
                     \si \cdot \sj
                     \left(\frac{R_0}{r_{ij}}\right)
                      F_{N}(x_{A}),\\
\label{(O_FN)}
\Omega_{F,N} &=& \sum_{i \ne j} \tau_{+}^{(i)} \tau_{+}^{(j)}
                     \left(\frac{R_0}{r_{ij}}\right) F_{N}(x_{A}),\\
\label{(GTT)}
\Omega_{GT'} &=&   \sum_{i \ne j} \tau_{+}^{(i)} \tau_{+}^{(j)}
                     \si \cdot \sj
                     \left(\frac{R_0}{r_{ij}}\right) F_{4}(x_{A}),\\
\Omega_{F'}  &=& \sum_{i \ne j} \tau_{+}^{(i)} \tau_{+}^{(j)}
                 \left(\frac{R_0}{r_{ij}}\right) F_{4}(x_{A}),\\
\label{(O_T)}
\Omega_{T'} &=& \sum_{i \ne j} \tau_{+}^{(i)} \tau_{+}^{(j)}
                     \big\{ 3 \sir \sjr -
                     \si \cdot \sj \big\}
                     \left(\frac{R_0}{r_{ij}}\right) F_{5}(x_{A}).
\ea
Here, $R_0$ is the nuclear radius, introduced to make the matrix
elements dimensionless (compensating factors have been absorbed into
the phase space integrals \cite{doi85}).
The following notations are used
$${\bfr}_{ij} = ({\overrightarrow r}_i -
                         {\overrightarrow r}_j ),\ \ \
r_{ij} = |{\bfr}_{ij}|, \ \ \ {\hat{\bfr}_{ij}^{~}}= {\bfr}_{ij}/r_{ij},\ \ \
x_A = m_A r_{ij}.$$

Here  ${\overrightarrow r}_i$ is the coordinate of the "ith" nucleon.
The above matrix elements have been written in the closure approximation,
which is well satisfied due to the large masses of the intermediate
particles.

The nucleon structure coefficients in \rf{Omega} are given by
\ba{alph}
\alpha^{(0)}_V &=& \left(\frac{ F_S^{(3)}}{f_A}\right)^2,
\alpha^{(0)}_A = -\left(\frac{ T_1^{(3)}}{f_A}\right)^2,\\
\alpha^{(1)}_V &=& -\left(\frac{m_A}{m_P}\right)^2 \alpha^{(0)}_A
\left[\frac{1}{4} + \left(\frac{T_2^{(3)}}{T_1^{(3)}}\right)^2 -
\frac{T_2^{(3)}}{T_1^{(3)}}\right],\\
\alpha^{(1)}_A &=& -\left(\frac{m_A}{m_P}\right)^2 \left[\alpha^{(0)}_A
\left(\frac{1}{6} - \frac{2}{3} \frac{T_2^{(3)}}{T_1^{(3)}} +
\frac{4}{3} \frac{T_3^{(3)}}{T_1^{(3)}}\right) +
\frac{1}{12} \left(\frac{F_P^{(3)}}{f_A}\right)^2\right],\\
\label{(alph_T)}
\alpha_T &=& -\left(\frac{m_A}{m_P}\right)^2 \left[\alpha^{(0)}_A
\left(\frac{1}{12} - \frac{1}{3} \frac{T_2^{(3)}}{T_1^{(3)}} + \frac{2}{3}
\frac{T_3^{(3)}}{T_1^{(3)}}\right) - \frac{1}{12}
\left(\frac{F_P^{(3)}}{f_A}\right)^2\right].
\ea
Here
$F_{S,P}^{(3)} \equiv F_{S,P}^{(3)}(0),\ T_{i}^{(3)} \equiv T_{i}^{(3)}(0)$.

Three different structure functions $F_i$ appear in eqs. \rf{O_i}-\rf{O_T}
(we use notations of ref. ~\cite{Vergados1}). They are
given by the integrals over the momentum ${\bf q}$ transferred between two
nucleons (see eq. \rf{NRIA})
\ba{integrals}
F_N (x_A) &=& 4 \pi m_{A}^{6}  r_{ij} \int \frac{d^3{\bf q}}{(2\pi)^3}
\frac{1}{(m_A^2 + {\bf q}^2)^4}
        e^{i {\bf q} {\bf r}_{ij}} \\
\label{(F_4)}
F_4 (x_A) &=& 4 \pi m_{A}^{4} r_{ij} \int \frac{d^3{\bf q}}{(2\pi)^3}
        \frac{{\bf q}^2}{(m_A^2 + {\bf q}^2)^4}
        e^{i {\bf q} {\bf r}_{ij}} \\
\label{(F_5)}
F_5 (x_A) &=& 2 \pi m_{A}^{4} r_{ij} \int \frac{d^3{\bf q}}{(2\pi)^3}
        \frac{{\bf q}^2 - \frac{1}{3}({\bf q}\cdot \hat{\bf r}_{ij})^2}
{(m_A^2 + {\bf q}^2)^4}
        e^{i {\bf q} {\bf r}_{ij}}.
\ea
These functions are analogous to the "neutrino potentials" for the
case of the light Majorana neutrino exchange. Nuclear energy denominators
typical for such momentum integrals can be neglected safely from
eqs. \rf{integrals}-\rf{F_5} in the case of heavy intermediate particles,
such as SUSY particles and the heavy Majorana neutrino $N$, with masses
much larger than the characteristic nuclear energy and are therefore
suppressed in \rf{integrals}-\rf{F_5}.
The analytic solutions of the integrals in eqs. \rf{integrals}-\rf{F_5} are
\ba{Fnn}
&F_N(x)  = \frac{x}{48} (3 + 3 x + x^2) e^{-x}, \ \
F_4(x)  = \frac{x}{48} (3 + 3 x - x^2) e^{-x},&\\ \nn
&F_5(x)  = \frac{x^3}{48} e^{-x}.&
\ea

At this stage we  point out that our formulas for
the coefficients \rf{alph}-\rf{alph_T} of the  transition
operators \rf{Omega}-\rf{Omega_f} disagree with
the corresponding formulas derived in
ref. \cite{Vergados1}. Particularly, in our case
$\alpha_A^{(0)} \leq 0$ while in ref. \cite{Vergados1}
this coefficient is positive. This sign difference has
an important consequence if one considers simultaneously
the supersymmetric and the heavy Majorana neutrino contributions.
It will be seen in the next section that
in both the  neutrino exchange and in the \rp~SUSY mechanisms
the dominant contributions correspond to the $\Omega_{GT,N}$
transition operator, initiating Gamow-Teller $0^+ \longrightarrow 0^+$
nuclear transitions. Comparing eq. \rf{Omega} with eq. \rf{Omega_N} one can see
that our formulas correspond to a constructive interference between
the dominant $\Omega_{GT,N}$ terms of these two mechanisms while
formulas from ref. \cite{Vergados1} correspond to a destructive one.
In the latter case both contributions can
cancel each other and by a proper choice of $\lambda'_{111}$ in
\rf{eta}-\rf{eta_end}
the  matrix element of $\znbb$-decay  can be set to zero at any values
of particle masses involved in the formulas. As a result the $\znbb$-decay
half-life limit would neither constrain supersymmetric
particle masses nor that for Majorana neutrinos.
Our formulas \rf{Omega}, \rf{Omega_f} and \rf{alph}-\rf{alph_T} always lead
to certain constraints on these masses. We return to a detailed discussion of
this point in section 6.

\section{Nuclear matrix elements in the
pn-QRPA approach}

While the formalism of the SUSY $\znbb$ decay outlined in the
previous sections is independent of the nuclear structure model used
to generate the wave functions, numerical values of matrix elements
are model dependent. We therefore  separate this part of
our work from the rest of the formalism in order to split up nuclear
and particle physics uncertainties in a most distinctive way.

\subsection{Basic summary of the nuclear model and its parameters}

Nuclear matrix elements have been calculated within pn-QRPA (proton
- neutron Quasiparticle Random Phase Approximation). pn-QRPA, the RPA for
charge-changing transitions has been developed by Halbleib and Sorenson
\cite{hal67} and during recent years widely been applied to double beta
decay calculations \cite{vog86}-~\cite{hir94}. In the present work we follow
essentially the description of \cite{mut89a}, \cite{mut89b}, \cite{hir94}.

Further we just briefly summarize the numerically most important
features.

We account for typically two major oscillator shells, for example
$3 \hbar \omega$ and $4 \hbar \omega$ in case of $^{76}$Ge. Single particle
energies are calculated from a Coulomb-corrected Woods-Saxon potential
with Bohr-Mottelson parameters \cite{boh69}, except that the strength of
the spin-orbit force has been reduced according to \cite{hir94} by
$\sim 9$ \% for a better agreement with experimental data. For
the pairing (in Bardeen-Cooper-Schriefer (BCS) approximation)
and RPA calculation we used consistently the Paris nucleon-nucleon
potential \cite{paris}. The strengths of the pairing interactions in the
BCS calculation have been adjusted to reproduce the experimentally observed
pairing gaps \cite{wap88}. In the RPA calculation, for the strength of
the particle-hole interaction a simple mass number dependence was used,
$g_{ph} = 1 + 0.002 A$ \cite{mut89a}. For the strength of the
particle-particle interaction numerical values of matrix elements are given
for values of $g_{pp}$, which have been fitted to experimentally
measured $\beta^+$/EC decay strengths \cite{mut89a}.

As mentioned in the previous sections, in the SUSY $\znbb$ decay the
virtually exchanged particles are assumed to be heavy. Thus, the
corresponding operators will be short-ranged. For this reason the
short-range part of the nuclear wave functions has to be treated carefully.
In addition to the finite nucleon size effects, which are
taken into account by the nucleon form factors in momentum space
(see eq. \rf{dip}), the nucleon-nucleon repulsion at short distances
must be accounted for. These short-range correlations are treated
by multiplying the two particle
wave functions  by the correlation function \cite{mil76}

\be{corr1}
 1 - f(r) = 1 - e^{-a r^2} (1 - b r^2).
\ee
The two parameters $a$ and $b$ can be related to each other,
so that effectively there is only one free parameter,
the so-called correlation length defined by

\be{corr2}
 l_c = - \int_{0}^{\infty} ds \big( [1+f(r)]^2-1 \big) \simeq {0.748\over
         {\sqrt{a}}},
\ee
with the standard value of $l_c = 0.7$ fm.

\subsection{Numerical results and discussion}
In the SUSY nuclear transition operators \rf{Omega}-\rf{Omega_f}
there appear 5 basic structures \rf{O_i}-\rf{O_T}.
These, combined with the nuclear wave functions give rise
to the following 5 nuclear structure matrix elements \footnote{Note,
that compared to the definitions of Fermi-type matrix elements for the
light neutrino case \cite{mut89b}, ~\cite{hir94} we took out the factor
$(f_V/f_A)^2$ here. In case of the SUSY mechanism of $\znbb$ decay all
appearing combinations of coupling constants have been absorbed
into the coefficients $\alpha^{(i)}$.}

\be{mgth}
{\cal M}_{GT,N} = \big< F |\Omega_{GT,N} | I \big>  =
\big< F | \sum_{i \ne j} \tau_{+}^{(i)} \tau_{+}^{(j)}
                     \si \cdot \sj
                     \left(\frac{R_0}{r_{ij}}\right)
                      F_{N}(x_{A})                | I \big>
\ee

\be{mfh}
{\cal M}_{F,N} = \big< F |\Omega_{F,N} | I \big> =
                 \big< F | \sum_{i \ne j} \tau_{+}^{(i)} \tau_{+}^{(j)}
                     \left(\frac{R_0}{r_{ij}}\right)
                      F_{N}(x_{A})                | I \big>
\ee

\be{mgtqq}
{\cal M}_{GT'} = \big< F |\Omega_{GT'} | I \big> =
\big< F | \sum_{i \ne j} \tau_{+}^{(i)} \tau_{+}^{(j)}
                     \si \cdot \sj
                     \left(\frac{R_0}{r_{ij}}\right)
                      F_{4}(x_{A})                | I \big>
\ee

\be{mfqq}
{\cal M}_{F'} = \big< F |\Omega_{F'} | I \big> =
                 \big< F | \sum_{i \ne j} \tau_{+}^{(i)} \tau_{+}^{(j)}
                 \left(\frac{R_0}{r_{ij}}\right)
                  F_{4}(x_{A})                | I \big>
\ee

\be{mtqq}
{\cal M}_{T'} = \big< F |\Omega_{T'} | I \big> =
                     \big< F | \sum_{i \ne j} \tau_{+}^{(i)} \tau_{+}^{(j)}
                     \big\{ 3 \sir \sjr -
                     \si \cdot \sj \big\}
                      \left(\frac{R_0}{r_{ij}}\right) F_{5}(x_{A})| I \big>
\ee
We have calculated these five basic matrix elements for the
experimentally most interesting isotopes. The results are given
in table 2. As expected ${\cal M}_{GT,N}$ and ${\cal M}_{F,N}$ have the
largest numerical values, whereas the other 3 matrix
elements give only minor corrections.
This can be simply understood, if one notes that
matrix elements ${\cal M}_{GT'}, {\cal M}_{F'}, {\cal M}_{T'}$
have additional $\bf{q}^2$
factors in the integrals \rf{F_4}-\rf{F_5} over the momentum $\bf{q}$,
transferred
between two decaying neutrons. The effective nuclear cut-off
for momentum transfer
is $q\ \lsim\ p_F$, where $p_F\sim 100$MeV is the momentum of the nucleon Fermi
motion inside the nucleus.
This then results in the relative suppression of the nuclear matrix
elements of the last three operators in eqs. \rf{mgtqq}-\rf{mtqq}
by factors of $(p_F/m_P)^2 \simeq$ (few) \%.

\begin{table}[t]
{Table 2: Nuclear matrix elements for SUSY $0\nu\beta\beta$ decay for
the experimentally most interesting isotopes calculated within
pn-QRPA. ${\cal M}_{\alpha}$ $\times 10^{x}$ implies that the matrix element
should be divided by $10^{x}$ to get the correct numerical value. These
factors are introduced here, just to show the differences in
magnitude of the different matrix elements.}\\[5mm]
\begin{tabular}{|c|c|c|c|c|c|}\hline
&&&&&\\
$^{A}$Y & ${\cal M}_{GT,N}$ $\times 10^{1}$ & ${\cal M}_{F,N}$ $\times 10^{2}$
&
${\cal M}_{GT'}$ $\times 10^{3}$ & ${\cal M}_{F'}$ $\times 10^{3}$  &
${\cal M}_{T'}$ $\times 10^{3}$ \\
&&&&&\\
\hline
&&&&&\\
$^{76}$Ge  & 1.13 & -4.07 & -7.70 & 3.06 & -3.09\\
&&&&&\\
\hline
&&&&&\\
$^{82}$Se  & 1.02 & -3.60 & -7.13 & 2.76 & -2.76\\
&&&&&\\
\hline
&&&&&\\
$^{100}$Mo & 1.29 & -4.89 & -7.88 & 3.52 & -4.93\\
&&&&&\\
\hline
&&&&&\\
$^{116}$Cd & 0.75 & -2.71 & -5.49 & 2.19 & -2.65\\
&&&&&\\
\hline
&&&&&\\
$^{128}$Te & 1.19 & -4.19 & -7.95 & 3.11 & -3.85\\
&&&&&\\
\hline
&&&&&\\
$^{130}$Te & 1.05 & -3.69 & -6.97 & 2.73 & -3.53\\
&&&&&\\
\hline
&&&&&\\
$^{136}$Xe & 0.58 & -2.03 & -3.81 & 1.49 & -1.81\\
&&&&&\\
\hline
&&&&&\\
$^{150}$Nd & 1.65 & -5.91 & -10.4 & 4.32 & -7.51\\
&&&&&\\
\hline
\end{tabular}
\end{table}

The data of table 2, however, shows also another interesting fact:
Corresponding matrix elements for different isotopes are relatively
similar to each other, with a spread of about a factor of $2$
from the mean only. As is known, much larger differences are usually
found in calculations of $\tnbb$ decay matrix elements as one goes from
one isotope to another. The similarity of the SUSY $\znbb$ decay
matrix elements is, on the other hand, not really a surprise.
Recall that due to the large masses of the intermediate particles
the transition operators are short-ranged.
Therefore only the part of the wave functions
at short distances contribute appreciably to the
matrix elements and nuclear structure effects, which are so
important in $\tnbb$ decay, play a less dominant role in SUSY
$\znbb$ decay. (See also the discussion in the next section.)

According to the eqs. \rf{Omega}, \rf{Omega_f} the 5 basic matrix elements
\rf{mgth}-\rf{mtqq} can then be combined  to define the following
nuclear matrix elements describing contributions to $\znbb$ decay
which correspond to {\it the} \rp SUSY mechanism
\ba{msq}
{\cal M}_{\tilde q} &=& \big< F |\Omega_{\tilde q} | I \big> = \\ \nn
             &=&{\frac{m_P}{ m_e}}
                \Big\{ \alpha_V^{(0)} {\cal M}_{F,N} + \alpha_A^{(0)} {\cal
M}_{GT,N}
               + \alpha_V^{(1)} {\cal M}_{F'} + \alpha_A^{(1)} {\cal M}_{GT'}
               + \alpha_T {\cal M}_{T'} \Big\}
\ea
\be{msl}
{\cal M}_{\tilde f} = \big< F |\Omega_{\tilde f} | I \big> =
                        {\cal M}_{\tilde q} |_{\forall T_i = 0}
\ee
and the conventional mass mechanism due to heavy Majorana neutrino exchange
\be{M_NN}
{\cal M}_N = \big< F |\Omega_N | I \big> =
{\frac{m_P}{ m_e}}\Big\{ \left(\frac{f_V}{f_A}\right)^2 {\cal M}_{F,N} -
                        {\cal M}_{GT,N} \Big\},
\ee
Table 3 shows  ${\cal M}_{\tilde q}$ and  ${\cal M}_{\tilde f}$
for the isotopes considered for the two sets of
input parameters for the
nucleon structure constants $\alpha^{(i)}$ as calculated from table 1.
\footnote{Although ${\cal M}_{\tilde q}$ and  ${\cal M}_{\tilde f}$
are negative numerically, positive values are given in table 3. This
was done to adjust the signs of matrix elements to the notation used
in our previous publications on $\znbb$ decay in left-right-symmetric
models \cite{mut89b}, \cite{hir94}.}

Again, it is noteworthy that both sets of parameters lead to rather
similar results. This is due to a partial cancellation of
the differences in the input values for the $\alpha^{(i)}$,
at least in the case of ${\cal M}_{\tilde q}$.

\begin{table}[t]
{Table 3: Nuclear matrix elements for SUSY $0\nu\beta\beta$ decay.
Shown are ${\cal M}_{\tilde q}$ and ${\cal M}_{\tilde f}$ for the two sets of
input values of coefficients for the $\alpha^{(i)}$ of table 1.}\\[5mm]
\begin{tabular}{|c|c|c|c|c|}\hline
&&&&\\
$^{A}$Y & A) ${\cal M}_{\tilde q}$ & B) ${\cal M}_{\tilde q}$ &
A) ${\cal M}_{\tilde f}$ & B) ${\cal M}_{\tilde f}$ \\
&&&&\\
\hline
&&&&\\
$^{76}$Ge  & 283 & 304 & 13.2 & 20.7 \\
&&&&\\
\hline
&&&&\\
$^{82}$Se  & 253 & 272 & 11.3 & 17.9 \\
&&&&\\
\hline
&&&&\\
$^{100}$Mo  & 328 & 356  & 23.6 & 33.5 \\
&&&&\\
\hline
&&&&\\
$^{116}$Cd  & 190 & 205 & 11.0 & 16.2 \\
&&&&\\
\hline
&&&&\\
$^{128}$Te  & 298 & 323 & 16.7 & 24.8 \\
&&&&\\
\hline
&&&&\\
$^{130}$Te  & 262  & 284 & 15.4 & 22.5 \\
&&&&\\
\hline
&&&&\\
$^{136}$Xe  & 143 & 155 & 7.91 & 11.8 \\
&&&&\\
\hline
&&&&\\
$^{150}$Nd  & 416 & 456 & 34.4 & 47.0 \\
&&&&\\
\hline
\end{tabular}
\end{table}

Table 3 moreover shows that the ${\cal M}_{\tilde f}$'s are much smaller
than the ${\cal M}_{\tilde q}$'s.
This difference can be simply understood because only scalar and pseudoscalar
currents $J_{S,P}$ contribute to ${\cal M}_{\tilde f}$,
but no tensor current $J_T$. Since in this case (see \rf{msl})
$\alpha_A^{(0)} \sim (T_1^{(3)})^2 = 0$, there is no contribution
from the large ${\cal M}_{GT,N}$ to ${\cal M}_{\tilde f}$ and consequently
${\cal M}_{\tilde f} \ll {\cal M}_{\tilde q}$. This result has important
implications in the numerical analysis of section 6.

\subsection{Uncertainties of the nuclear structure
matrix elements}

We have investigated the dependence of the numerical values of the
matrix elements on our choice of nuclear model parameters.
Besides the numerical uncertainties discussed here, there will also be
deviations from the "true" matrix element due to model approximations.
These, however, can not be quantified exactly. Some confidence in
the model might be derived from the fact that the $\tnbb$ decay
half-life of $^{76}$Ge \cite{hdmo94a} has been {\it predicted}
correctly \cite{mut89a} within a factor of $2$ (the $\tnbb$ decay
matrix element within $\sqrt{2}$).

In the following the three
most important parameters are discussed in some detail. Other
parameters, like for example $g_{ph}$ or the use of another
nucleon-nucleon potential (Reid soft-core potential \cite{reid})
instead of the Paris force have been found to have negligible
effects.

In calculations of $\tnbb$ decay \cite{vog86}, ~\cite{civ87}, ~\cite{mut89a}
it has been shown that
the $\tnbb$ decay matrix elements calculated within pn-QRPA do
strongly depend on the strength of the particle-particle
interaction $g_{pp}$. We have therefore calculated SUSY $\znbb$
decay matrix elements as a function of $g_{pp}$ for all isotopes
considered in the present work. A typical example is shown in
fig. 3. Compared to the results of $\tnbb$ decay calculations
only a very weak dependence of the matrix elements on $g_{pp}$ is
found. A variation of $g_{pp}$ within $\pm 2$ standard deviations
from its fitted value \cite{mut89a} changes the SUSY matrix
elements only by about $10$ \%.

The explanation of this weak sensitivity of matrix elements on
$g_{pp}$ is similar to
the one discussed in $\znbb$ decay
calculations for light neutrino exchange \cite{mut89b}.
Due to the selection
rules for the Gamow-Teller operator, in $\tnbb$ decay the intermediate
nuclear states are $1^+$-states, exclusively.
On the other hand, also higher multipoles contribute to the $\znbb$
matrix elements in eqs. \rf{mgth}-\rf{mtqq} similarly
to the above mentioned case of the light neutrino exchange mechanism.
A typical example of a multipole decomposition for ${\cal M}_{GT,N}$
and ${\cal M}_{F,N}$
is shown in fig. 4. Quite large contributions up to high
multipoles are found. Since the particle-particle force mainly
affects the $1^+$-states \cite{mut89b}, SUSY $\znbb$ decay matrix
elements are not very sensitive to the actual choice of $g_{pp}$.

Figure 5 shows the nuclear matrix elements for the example of $^{130}$Te
as a function of the momentum cutoff factor $m_A$. It is seen that
${\cal M}_{GT,N}$, ${\cal M}_{F,N}$ and ${\cal M}_{T'}$ are rather stable
against
variations of $m_A$, while ${\cal M}_{GT'}$ and ${\cal M}_{F'}$
are more sensitive.
Since ${\cal M}_{\tilde q}$ is dominated by ${\cal M}_{GT,N}$,
it will not be very sensitive to variations of $m_A$ as well.
${\cal M}_{\tilde f}$, on the
other hand, has to be expected to depend more strongly on $m_A$.

Figure 6 then plots matrix elements as a function of the correlation
length $l_c$ for the example of $^{150}$Nd.
As can be read off, a $20$ \% change in $l_c$ typically
changes calculated ${\cal M}_{GT,N}$ and ${\cal M}_{F,N}$
by about ($20-30$) \%.
Similar to the case when the cutoff factor is varied, ${\cal M}_{GT'}$ and
${\cal M}_{F'}$ are found to be more sensitive to changes in $l_c$.

Since changes in all three parameters can lead to both larger and smaller
matrix elements, the influence of a simultaneous change of these can not
be estimated by simply adding up the individual errors. To estimate the
total uncertainty, we have therefore calculated ${\cal M}_{\tilde q}$ and
${\cal M}_{\tilde f}$ at 125 grid points in this 3-dimensional parameter
space.
(Parameter variations within $\pm 20$ \% of their standard values
for $l_c$ and $m_A$ and within $2$ standard deviations for $g_{pp}$.)
{}From this set we then calculated the "mean" matrix elements, the
corresponding standard deviations and extreme values.

In case of ${\cal M}_{\tilde q}$ these mean matrix elements are always very
similar to those calculated for the standard values of input parameters
(typical differences up to a few \%).
The standard deviations for ${\cal M}_{\tilde q}$
and its extreme values deviate typically less than $20$ \% and
$50$ \%, respectively, from the mean.

The situation is different in case of ${\cal M}_{\tilde f}$.
As discussed above,
${\cal M}_{GT'}$ shows a larger sensitivity to model parameter variations than
do other matrix elements. Moreover, in case of ${\cal M}_{\tilde f}$ there
occurs a destructive interference between contributions from ${\cal M}_{F,N}$,
${\cal M}_{T'}$ and ${\cal M}_{GT'}$. In certain regions of the model parameter
space ${\cal M}_{GT'}$ tends to cancel the contributions from the other two
matrix elements. The conclusion therefore unfortunately must be that
${\cal M}_{\tilde f}$ has to be considered to carry a large uncertainty:
At extreme combinations of parameters ${\cal M}_{\tilde f}$ is smaller than
(larger than) in the standard calculation by up to a factor of $5$ ($2$).

To summarize this discussion, it can be stated that while ${\cal M}_{\tilde q}$
can be reliably calculated, the numerical value of
${\cal M}_{\tilde f}$ must be considered to be rather uncertain.
Fortunately we will see in the next section that only the
${\cal M}_{\tilde q}$  matrix element is relevant for the extraction of
the \rp MSSM parameters from
$\znbb$ experimental data on $\znbb$ decay.

\section{Constraints on the \rpt ~MSSM parameter space}

In the following we will analyse constraints on the \rp~MSSM
parameter space using the experimental half-life limit of $^{76}$Ge,
recently measured by the Heidelberg-Moscow collaboration
\cite{hdmo94} (for constraints derived from experiments on other isotopes,
see below),

\be{expt12}
T_{1/2}^{\znbb}(^{76}Ge, 0^+ \rightarrow 0^+) \hskip2mm \big> \hskip2mm
5.6 \times 10^{24} years \hskip10mm 90 \% \ c.l.
\ee

The theoretical expression for the inverse $\znbb$ half-life (see \rf{tot})
corresponding to the reaction matrix element ${\cal R}_{\znbb}$
in eq. \rf{R_0nu} can be written as,

\be{tht12}
\big[ T_{1/2}^{\znbb}(0^+ \rightarrow 0^+) \big]^{-1} =
 G_{01} \Big\{ (\eta_{\tilde g} + \eta_{\chi}) {\cal M}_{\tilde q}
 + ( \eta_{\chi {\tilde e}} + \eta'_{\tilde g}- \eta_{\chi \tilde f})
     {\cal M}_{\tilde f} + \eta_N {\cal M}_N\Big\}^2.
\ee
In the numerical analysis
we will employ for definiteness the matrix
elements for Set A of the coefficients $\alpha^{(i)}$ (see table 1).
Matrix elements of Set B would only yield slightly more stringent bounds,
but not change any of the arguments presented below.

Neglecting for the moment the contribution from Majorana neutrino exchange
(the last term in eq. \rf{tht12}) we will first analyse the supersymmetric
contribution alone.

Eqs. \rf{expt12}, \rf{tht12} define an excluded area within
a 5-dimensional \rp ~MSSM parameter space
\be{param_space}
\{\tan\beta, \mu, m_{\tilde g}, m_0, \lambda'_{111}\}.
\ee
These parameters, as explained in section 2, completely define
the supersymmetric particle spectrum and their interactions initiating
$\znbb$.
However,
there are  too many free parameters to be
defined from only one constraint (eq. \rf{expt12}). Fortunately,
the analysis gets considerably simplified realizing that the
contributions from the five different lepton number violating parameters
$\eta_{\tilde g},\eta_{\chi}, \eta_{\chi\tilde e},
\eta_{\chi \tilde f}, \eta'_{\tilde g}$
enter eq. \rf{tht12} with very different magnitudes. It can be seen
from eqs. \rf{eta}-~\rf{eta_end} that
\be{domin}
\eta_{\tilde g}, \eta'_{\tilde g} >> \eta_{\chi}, \eta_{\chi\tilde e},
\eta_{\chi \tilde f}\ \ \
\mbox{if}\ \ \ m_{\tilde q} \sim m_{\tilde e},
\  m_{\chi_i}\gsim 0.02 m_{\tilde g}
\ee
with the values \rf{Coup} of the gauge coupling constants
$\alpha_2$ and $\alpha_s$ and for any field composition of
the neutralino states $\chi_i$ (see eq. \rf{admix}).
The last mass inequality in \rf{domin} is well satisfied even for
a gluino mass
$m_{\tilde g}$ as large as 500 GeV if  $m_{\chi_i} \gsim 20$ GeV. The latter
is guaranteed by the present experimental lower bound on the lightest
neutralino $\chi \equiv \chi_1$ from the LEP experiments ~\cite{lep_limit}
$m_{\chi}\gsim 20$GeV. Later on we consider the gluino dominance more
carefully and will see that it holds for $m_{\tilde g}$ up to 1 TeV.
Another interesting fact can be derived from the nuclear matrix elements
${\cal M}_{\tilde q, \tilde f}$ of table 3.
It can be seen that the value of
${\cal M}_{\tilde q}$ is much larger than the one of ${\cal M}_{\tilde f}$.
Therefore, taking into account gluino dominance eq. \rf{domin}, it follows
that the combination $\eta_{\tilde g}{\cal M}_{\tilde q}$ absolutely dominates
in the half-life formula eq. \rf{tht12}. The advantage of this fact
is twofold. First, only three \rp MSSM parameters
$m_{\tilde q}, m_{\tilde g}$ and $\lambda'_{111}$
are involved in the numerical analysis.
Second, as stated at the end of section 5, the nuclear model
calculations for the matrix element ${\cal M}_{\tilde q}$
are much more reliable than those for ${\cal M}_{\tilde f}$.
Thus, we are lucky to have a theoretically well controlled
nuclear structure dependence in the dominant term of eq. \rf{tht12}.
Having this in mind we proceed with the numerical analysis.

Two limiting cases are interesting to discuss:
a) $m_{{\tilde d}_R} = m_{{\tilde u}_L} \equiv
m_{\tilde q}$ and b) $m_{{\tilde u}_L} \gg m_{{\tilde d}_R}
\equiv m_{\tilde q}$ (the case
$m_{{\tilde d}_R} \gg m_{{\tilde u}_L}$ is equivalent).
The following bounds are obtained

\be{fude}
Case \hskip3mm a) \hskip10mm
\lambda'_{111} \le 3.9 \times 10^{-4} \Big({m_{\tilde q}\over{100 GeV}} \Big)^2
 \Big({m_{\tilde g}\over{100 GeV}} \Big)^{(1/2)} ,
\ee

\bigskip
\be{fude2}
Case \hskip3mm b) \hskip10mm
\lambda'_{111} \le 5.6 \times 10^{-4} \Big({m_{\tilde q}\over{100 GeV}} \Big)^2
 \Big({m_{\tilde g}\over{100 GeV}} \Big)^{(1/2)} .
\ee
\bigskip
These two extreme cases differ just by a factor of $\sqrt{2}$, and for
all other ratios of ($m_{{\tilde d}_R} / m_{{\tilde u}_L}$) limits
between case a) and case b) are found. Motivated by the MSSM mass
formulas, eq. \rf{SfermionMass}-\rf{smass_end},
in the following discussion we will
always assume $m_{{\tilde d}_R} = m_{{\tilde u}_L}$ which corresponds
to the case a).

A graphical representation of this bound (eq. \rf{fude})
is shown in fig. 7. The area on
the backside of the surface is forbidden by the $\znbb$ decay constraint
\rf{expt12}. Shown is the mass range between 10 GeV and 1 Tev
and a coupling range between $10^{-3}$ and $2 \times 10^{-1}$.
It is interesting to note that employing the upper bounds
$m_{\tilde q, \tilde g} \lsim 1 $TeV, motivated by the SUSY naturalness
argument, one can obtain from \rf{fude} an upper bound for
the \rp Yukawa coupling constant
$$ \lambda'_{111} \lsim 0.124. $$

As mentioned before we have analysed, how the presence of a non-zero
neutralino contribution to eq. \rf{tht12}, proportional to
$ \eta_{\chi}$,  \footnote{Although in general
all four neutralino mass eigenstates contribute to the $\znbb$ decay
rate, it is sufficient to consider only the lightest neutralino
with mass $m_{\chi}$. } affects the bounds derived from eq. \rf{fude}.
{}From eq. \rf{tht12}, one can expect that bounds derived from
$\eta_{\tilde g}$
would be sharpened when $\eta_{\chi}$ is switched on. The large
difference in magnitude, however, leads to the result that only for
very light neutralinos this effect will be of importance. This is
displayed in fig. 8, where the excluded area in the ($m_{{\tilde q}}$,
$m_{{\tilde g}}$, $m_{{\chi}}$)-space is shown for the example of
$\lambda'_{111} = 10^{-2}$. For a neutralino heavier than a few GeV, the squark
mass and gluino mass bounds get independent of the actual value
of $m_{\chi}$. As can be shown, this is true for {\it any} neutralino
composition. Since such a light neutralino mass eigenstate is already
excluded by the experiments at LEP \cite{lep_limit}, which tell us
that $m_{\chi} \ge 20$ GeV, the conclusion is that the limits of fig. 7
are practically independent of the neutralino mass.

Now another important  question arises. How much will the obtained limits
on superparticle masses and the \rp Yukawa coupling
be affected by uncertainties of the nuclear structure calculation?
{}From the definition of the lepton number violating parameters $\eta$
eqs. \rf{eta}-(~\ref{(eta_end)}) and the $\znbb$  decay constraint
eqs. \rf{expt12}-\rf{tht12}  the following conclusion can be made.
Any shift of the nuclear matrix element ${\cal M}_{\tilde q}$
by the value $\Delta {\cal M}_{\tilde q}$, associated with
the theoretical uncertainties, would change the extracted limits
on $\lambda'_{111}$, gaugino $m_{\tilde g, \chi}$ and sfermion
$m_{\tilde q, \tilde e}$ masses as follows

\ba{unc}
 \frac{\Delta m_{\tilde g, \chi}}
           {m_{\tilde g, \chi}} &\sim&
\frac{\Delta {\cal M}_{\tilde q}}{{\cal M}_{\tilde q}},\\
\frac{\Delta \lambda'_{111}}{\lambda'_{111}} &\sim&
1\  - \  \big(1 +  \frac{\Delta {\cal M}_{\tilde q}}
{{\cal M}_{\tilde q}} \big)^{-\frac 1 2},\\
\frac{\Delta m_{\tilde q}}{m_{\tilde q}} &\sim&
\big(1 +  \frac{\Delta {\cal M}_{\tilde q}}
{{\cal M}_{\tilde q}} \big)^{\frac 1 4} - 1.
\ea

{}From these equations one can see that the limits on
squark masses $m_{\tilde q}$
and the \rp Yukawa coupling constant
$\lambda'_{111}$, deduced from $\znbb$ decay
depend only very weakly on the nuclear physics uncertainties.
For instance, a change of ${\cal M}_{\tilde q}$ by even a factor of $2$
changes the squark mass limit by less then $20$ \%.

Up to now, we have not considered the influence of a possible contribution
from Majorana neutrinos to $\znbb$ decay.
In this paper we bound ourselves to the heavy Majorana neutrino (N)
contribution, as discussed at the end of sect. 2.
In section 4 it was shown that contributions to $\znbb$ decay
from the SUSY mechanism and the ordinary neutrino mass mechanism
add up coherently. This is opposite to the results of \cite{Vergados1},
where a destructive interference was derived.
The combined constraint
in the ($m_{\tilde q}$,$m_{\tilde g}$,$\langle m_N \rangle$)-space,
where $\langle m_N \rangle$ is the {\it effective} heavy
Majorana neutrino mass
(see eq. \rf{m_N}),
is shown in fig. 9, for the case of
$\lambda'_{111}= 10^{-2}$. It can be seen that limits on $\langle m_N \rangle$
remain valid also, if one allows contributions from supersymmetric
theories to $\znbb$ decay
(or, vice versa $\znbb$ decay limits on the \rp MSSM
are not sensitive to the actual value of the neutrino mass). This is an
important result of the present work, since for a destructive
interference
between the neutrino mass and the SUSY mechanisms of $\znbb$ decay,
it would always
be possible to find regions in parameter space, where {\it no}
constraints from $\znbb$ decay could be derived.
As an interesting by-product of this analysis, we note that the
current limit \rf{expt12} on the $\znbb$ decay half-life of $^{76}$Ge
implies
\be{nulim}
\langle m_N \rangle \hskip3mm \gsim \hskip3mm
5.1 \times 10^{7} \hskip3mm \mbox{GeV}.
\ee

As mentioned above, currently $^{76}$Ge provides the most stringent
limits on $\znbb$ decay. For completeness, we have calculated
the limits on supersymmetric parameters in the form of eq. \rf{fude},
for all isotopes considered in the present work. A summary of these
limits, together with the references for the experimental
half-life limits, are given in table 4.

\begin{table}[t]
{Table 4: Comparison of limits on supersymmetric parameters derived
from different $\beta\beta$ decay experiments
(see quoted references for half-life limits),
written in the form of eq.
\rf{fude}: \\
$\lambda'_{111} \le \epsilon \Big({m_{\tilde q}\over{100 GeV}} \Big)^2
 \Big({m_{\tilde g}\over{100 GeV}} \Big)^{(1/2)}$. \\
\\
Currently
$^{76}$Ge provides the most stringent limits. ( [Ref.] $^a$ 90 \% c.l.,
whereas [Ref.] $^{b}$ 68 \% c.l. only.)}\\[5mm]
\begin{tabular}{|c|c|c|c|c|c|}\hline
&&&&&\\
$^{A}$Y & $\epsilon$ & Ref. &
$^{A}$Y & $\epsilon$ & Ref. \\
&&&&&\\
\hline
&&&&&\\
$^{76}$Ge  & $3.9 \times 10^{-4}$ & \cite{hdmo94} $^{a}$ &
$^{128}$Te  & $4.9 \times 10^{-4}$ & \cite{ber92} $^{b}$ \\
&&&&&\\
\hline
&&&&&\\
$^{82}$Se  & $1.1 \times 10^{-3}$ & \cite{ell92} $^{b}$ &
$^{130}$Te  & $1.1 \times 10^{-3}$ & \cite{ale94}$^{a}$  \\
&&&&&\\
\hline
&&&&&\\
$^{100}$Mo  & $7.5 \times 10^{-4}$ & \cite{als93} $^{b}$ &
$^{136}$Xe  & $6.8 \times 10^{-4}$ & \cite{vui93} $^{a}$ \\
&&&&&\\
\hline
&&&&&\\
$^{116}$Cd  & $2.5 \times 10^{-3}$ & \cite{eji94} $^{a}$ &
$^{150}$Nd & $9.7 \times 10^{-4}$ & \cite{moe94}  $^{a}$  \\
&&&&&\\
\hline
\end{tabular}
\end{table}

\section{Comparison of constraints from $\znbb$ decay with other
experiments}
It is interesting, to compare limits on the \rp MSSM parameters
from $\znbb$ decay, with those derived from other experiments.
These constraints can be derived from low-energy processes involving
virtual superparticles \cite{bar89} or from direct accelerator searches
Since $\znbb$ decay is sensitive only to the first generation lepton
number violating \rp coupling, we will restrict the discussion to
limits on $\lambda'_{111}$. Limits on other couplings might be found in
the literature~\cite{roy92}, ~\cite{bar89}.

In ref.  \cite{bar89} various low-energy processes have been analysed.
It was concluded that the most restrictive limit on $\lambda'_{111}$ might be
derived from charged current universality. The limitation follows
from the fact that the  existence of \rp Yukawa coupling
$\lambda'_{ijk} L_i Q_j\bar{D}_k$ gives an extra contribution
to quark semileptonic decays (e.g., in nuclear $\beta$ decay).
The effective four-fermion interaction induced by the \rp MSSM contribution
in fig. 10 has a $(V - A)\otimes(V - A)$ form identical
to the one derived in the standard model. Therefore
its contribution is equivalent to a shift in the Fermi constant
$G_F$.
If one assumes that only one \rp operator has a sizable
coupling constant, for instance $\lambda'_{111}L_1 Q_1 \bar{D}_1$
to a violation of charged current (CC) universality ~\cite{bar89} because
the shift of $G_F$ is different for different generations.
The experimental limits impose the following bound at
the two sigma level

\be{fude4}
\lambda'_{111} \le 0.03  \Big({m_{\tilde d_R}\over{100 GeV}} \Big).
\ee
If one allows more than one \rp operator to contribute the violation of
CC universality is reduced and the above bound is weakened.
   Comparing \rf{fude} and \rf{fude4} one can see that for masses in
the range of $100$ GeV, the bound \rf{fude4} is less restrictive
by nearly 2 orders of magnitude than the bound \rf{fude} derived
from $\znbb$ decay.

Accelerator searches on supersymmetric particles are usually based on
the assumption of R-parity conservation. In this case the LSP,
which is assumed to be the neutralino $\chi$,
is stable and only weakly interacting.
Therefore it escapes from the detector yielding the prominent
missing transverse energy (\et) signature of SUSY events. The CDF bounds
on superparticle masses ~\cite{CDF} rely essentially on \et  signals
and are not valid in the \rp case.
However, in ref. \cite{roy92} it has been shown that limits
on superparticle masses in the \rp case can be derived using
the CDF dilepton search data. The corresponding mass limit is
\be{CDF}
m_{{\tilde g},{\tilde q}} \ge 100 \mbox{GeV},
\ee
independently of $\lambda'_{111}$ if
$\lambda'_{111} \ge 10^{-5}$ is assumed.
For smaller values of $\lambda'_{111}$, the LSP
has a negligible decay probability
inside the detector and the \rp signal has \et pattern as in
the $R_p$ conserving case.
Then the limits obtained in ref. ~\cite{CDF} can
be applied for such small \rp couplings.
A recent calculation  \cite{bae94} shows that even
larger masses than in eq. \rf{CDF} might be probed at
the TEVATRON in the near future.

There are in the literature other proposals for searching  \rp SUSY
signals in future accelerator experiments.

The isolated like-sign dilepton signature
is
proposed as
a characteristic feature of the \rp events at the LHC

Recently searching for \rp SUSY events in deep inelastic ep-scattering
experiments with the ZEUS detector at HERA has been proposed \cite{but93}.
Two sets of signals could be identified with these events.
They are the \rp resonant squark production followed by the \rp cascade decay
of the neutralino and the MSSM $R_p$ conserving production of
selectron and squark with their subsequent \rp decay to ordinary matter.
It was advocated that searching for these signals provide HERA with a
promising discovery potential for \rp SUSY.

A comparison of the bounds, which according to \cite{but93} might be reached
with one year of HERA data, and the other limits discussed above
with the limits we have obtained from the $\znbb$ decay constraint \rf{expt12}
is shown in  fig. 10 in the  $\lambda'_{111}-m_{\tilde q}$ plane.
In the case of $\znbb$ decay,
limits for 2 different values
of the gluino mass are shown. It can be seen, that even for a
gluino mass as large
as 1 TeV which is marginal from the point of view of SUSY naturalness,
the present double beta decay half-life limit yields the most restrictive
bounds.
\footnote{This result, however, does not touch upon other \rp
couplings than $\lambda'_{111}$ which are unaccesible for $\znbb$,
but can be probed in other processes.}

To summarize this discussion, $\znbb$ decay
allows to stringently restrict $R_p$ violating supersymmetric theories.
We have shown that
these limits are more stringent than those from other low-energy processes
as well as those which can be derived from
the HERA experiments (see fig. 10).

\section{Conclusions}

   We have investigated contributions to $\znbb$ decay from supersymmetric
theories with explicit R-parity violation. The complete set of
diagrams describing quark-lepton interactions at short-distances
has been considered.
On this basis we have obtained the relevant low-energy effective Lagrangian
in terms of nucleon and lepton currents. Then we have derived
the transition operators describing
$0^+\rightarrow 0^+$ nuclear transitions induced by the \rp MSSM interactions.

   It has been found that contributions from the \rp SUSY mechanism of
$\znbb$ decay add coherently to the well-known neutrino mass mechanism.
As a result,  limits on the Majorana neutrino mass derived from $\znbb$
decay  remain valid also if the supersymmetric contributions are
taken into account.

   We have calculated nuclear matrix elements of these transition operators
within a realistic nuclear structure model for experimentally interesting
isotopes.

   Special attention has been paid to the theoretical uncertainties
of the nuclear matrix elements and to the question how these uncertainties
affects limits on the \rp MSSM model parameters we extracted from $\znbb$.
We were able to conclude that these limits do only very weakly depend on
nuclear physics uncertainties.

   Using existing experimental lower bounds on the $\znbb$ decay
half-life $T_{1/2}$,
we then analysed constraints on
the \rp MSSM parameters from $\znbb$ decay.
The most restrictive limits are found from the current $^{76}$Ge $\znbb$ decay
experiment by the Heidelberg-Moscow collaboration.
We conclude that $\znbb$ decay imposes very restrictive bounds on
the lepton number violating sector of  \rp SUSY models.
We have presented these bounds as a 3-dimensional exclusion plot
in the space of the \rp Yukawa coupling constant
$\lambda'_{111}$, squark $m_{\tilde q}$ and gluino $m_{\tilde q}$
masses (fig.7) as well as a 2-dimensional one in the
$\lambda'_{111}$-$m_{\tilde q}$ plane (fig.10) which also shows bounds
from other low and high energy processes.

   We infer that the $\znbb$ bounds  are able to compete with or
are even more stringent than those derived from current and
near future accelerator experiments.
Particularly, they exclude the domain which is accessible
for the experiments with the ZEUS detector at HERA.

\bigskip
\centerline{\bf ACKNOWLEDGMENTS}

\bigskip
We thank V.A.~Bednyakov, V.B.~Brudanin,
M.~Lindner and
J.W.F.~Valle  for helpful discussions.
The research described in this publication was made possible in part by
Grant No.RFM000 from the international Science Foundation.

\section{Appendix}

Here we define some notations used in the main text. A complete set of
the relevant formulas can be found in ref. ~\cite{doi85}.

The $\znbb$ reaction matrix element  ${\cal R}_{\znbb}$ is defined as follows
\ba{R}
\big <F| S - 1 |I \big> =
                        2\pi i \delta(\epsilon_1 + \epsilon_2 + E_f - E_i)
                        {\cal R}_{\znbb}
\ea

where $E_i, E_f$, $\epsilon_{1,2} = \sqrt{\vec{p}_{1,2}^{2} + m_{e}^{2}}$
are the energies of the initial and final nuclear states
as well as the energies of two outgoing electrons with
the 3-momenta ${\vec p}_i$.

For the Lagrangian represented in the form
\ba{G_L}
{\cal L} = \frac{G_{F}^{2}}{2} m_{P}^{-1}
\left\{\bar{e} (1 + \gamma_5) e^{\bf c} \right\} \sum_{i} \eta_i J_i J_i,
\ea
the corresponding reaction matrix element can be written as
\ba{matr}
{\cal R}_{\znbb} = \frac{G_{F}^{2}}{\sqrt{2}} m_{P}^{-1} C_{0\nu}^{-1}
\left\{\bar{e} (1 + \gamma_5) e^{\bf c} \right\} \sum_{i} \eta_i
\Big<F| \Omega_i |I \Big>
\ea
The only approximation made here is that nuclear matrix elements
are independent of the final-state electron energies $\epsilon_{1,2}$
and momenta ${\vec p}_{1,2}$.
This approximation is well satisfied in the $\znbb$ decay, because
$\epsilon_{1,2}\lsim T_0\sim (1-3)$MeV ($T_0$ is the energy release)
while the typical energy scale of the nuclear matrix elements is given
by the nucleon Fermi momentum, $p_F \sim 100$MeV.
Corrections to eq. \rf{matr} are smaller than 1 \%.

We introduce transition operators $\Omega_i$ as
\ba{O_def}
\big <F|\Omega_i |I \big> = \frac{C_{0\nu}}{(2\pi)^3}
                            \int d^3{\bf q} F_{i}^{2}({\bf q}^2)
                            \int d^3{\bf x} d^3{\bf y}
                            e^{i {\bf q}({\bf x} - {\bf y})}
\big <F|J_i({\bf x}) J_i({\bf y}) |I \big>.
\ea
The reaction matrix element \rf{matr} does not depend on the value of
the numerical coefficient $C_{0\nu}$, which is introduced to bring
the normalization of $\Omega$ in a coincidence with the literature
\be{C_0nu}
 C_{0\nu} = 4 \pi \frac{m_P}{m_e} \frac{R_0}{f_{A}^{2}} m_A^{-2}
\ee
The formula \rf{O_def} is written for the case of
heavy intermediate particles,
in the closure approximation and for the outgoing electrons
in S-wave states.

It is a common practice to use the
non-relativistic impulse approximation (NRIA)
for nuclear matrix element calculations. In this approximation the transition
operator $\Omega_i$, describing $N_a N_b \longrightarrow P_a P_b$ transitions
of two initial neutrons into two final protons induced by the current
$J_i$,  takes the form

\ba{NRIA}
\Omega_i &=& \frac{C_{0\nu}}{(2\pi)^3}
                     \sum_{a\neq b} \int d^3{\bf q} F_{i}^{2}({\bf q}^2)
                            e^{i {\bf q}\cdot {\bf r}_{ab}} \times \\ \nn
&&\times  <P_a({\bf p}_a) |J_i|N_a({\bf p'}_a)>_{nr}
<N_b({\bf p}_b) |J_i|N_b({\bf p'}_b)>_{nr}.
\ea
Here ${\bf q} = {\bf p}_a - {\bf p}'_a = {\bf p}'_b - {\bf p}_b$
is the momentum transfer,
$F_{i}({\bf q}^2)$ are the nucleon form factors and
the subscript $| >_{nr}$ means non-relativistic limit for the
corresponding nucleon matrix element. The summation over all pairs $N_a N_b$
of initial neutrons is implied.
${\bf r}_{ab}$ is the separation between these two neutrons.

The $\znbb$ observables can be calculated on the basis of
the reaction matrix element ${\cal R}_{\znbb}$.

The differential width of $\znbb$ is
\ba{dif}
d\Gamma = 2\pi \delta(\epsilon_1 + \epsilon_2 + E_f - E_i)
          \overline{|{\cal R}_{\znbb}|^2} d\Omega_1 d\Omega_2,
\ea
the phase space factors are
\ba{Phase}
d\Omega_i = \frac{d^3\vec{p}_i}{(2\pi)^3}.
\ea

The $\znbb$ half-life formula can be written as
\ba{tot}
\left[T_{1/2}^{\znbb}\right]^{-1} = \overline{|{\cal R}_{\znbb}|^2} G_{01}
\ea
where $G_{01}$ is the leptonic phase space integral, calculated
according to the prescription of \cite{doi85} and has the form

\ba{G_01}
 G_{01} = {a_{0\nu}\over{(m_e R_0)^2 ln(2)}}
          \int d\Omega_1 d\Omega_2 b_{01}.
\ea
Here, $a_{0\nu}=(G_F cos(\Theta_C) f_A/f_V)^4 m_e^9/(64 \pi^5 \hbar)$
involving only physical constants, $R_0$ is the nuclear
radius.
The kinematical factor $b_{01}$ accounts for the Coulomb distortion
of the electron waves
and can be taken from ref. ~\cite{doi85}.


{\large\bf Figure Captions}\\

\begin{itemize}

\item[Fig.1a] Basic diagram for neutrinoless double beta decay.

\item[Fig.1b] Basic diagram for neutrinoless double beta decay at
the quark level.

\item[Fig.2a] Feynman graphs for the conventional mechanism
of $\znbb$ decay by massive Majorana neutrino exchange.

\item[Fig.2b] Feynman graphs for the supersymmetric contributions
to $\znbb$ decay.

\item[Fig.2c] New "non-diagonal" Feynman graphs for supersymmetric
contributions to $\znbb$ decay \cite{HKK}.

\item[Fig.3a] SUSY $\znbb$ decay matrix elements as a function of the
particle-particle strength $g_{pp}$, for the example of $^{76}$Ge.
Shown are $M_{GT,N}$ (full line) and $M_{F,N}$ (dashed line).

\item[Fig.3b] As fig. 3.a, but for
${\cal M}_{GT'}$ (full line) and ${\cal M}_{F'}$ (dashed line)
and ${\cal M}_{T'}$
(dash-dotted line).

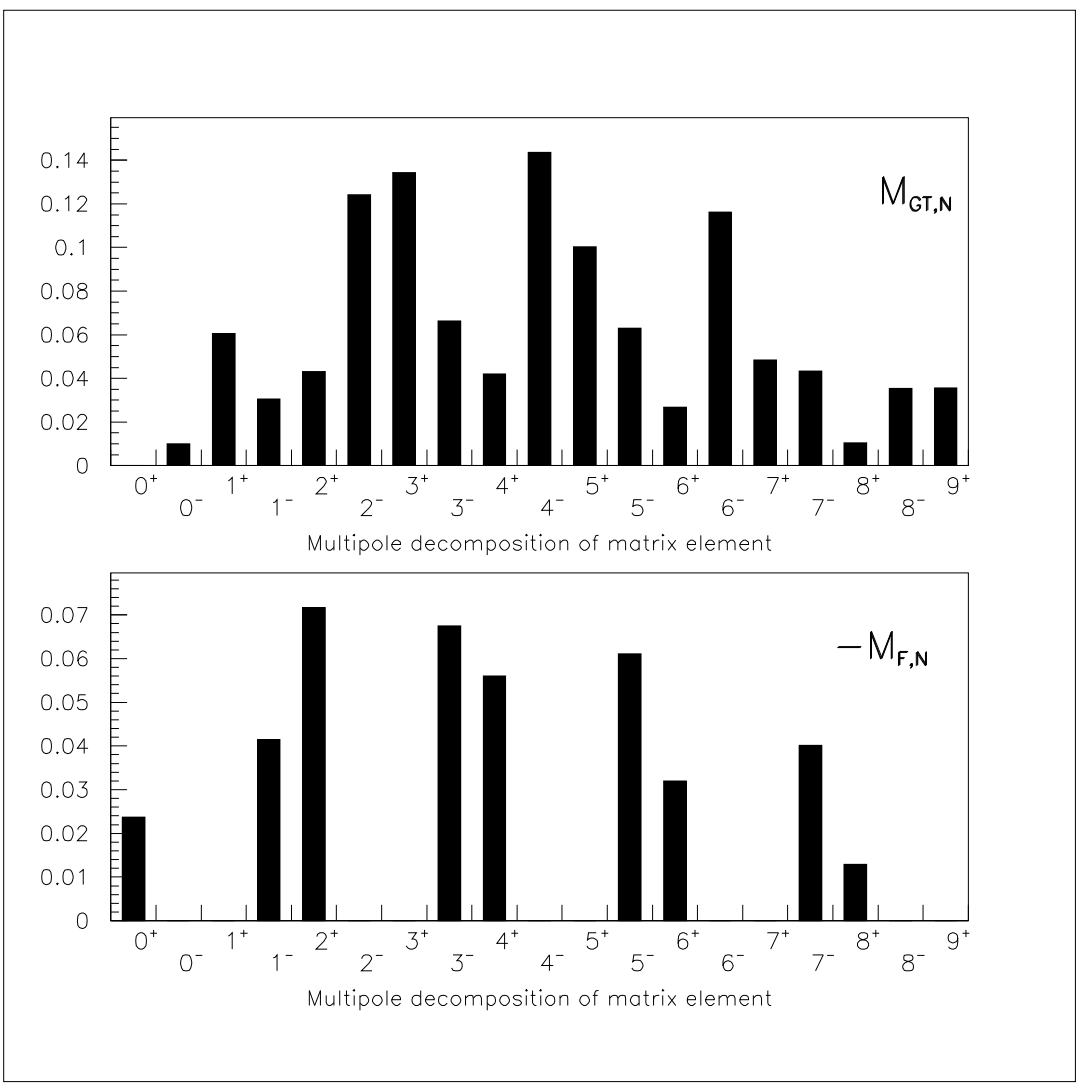
\item[Fig.4] Multipole decomposition of the matrix elements
${\cal M}_{GT,N}$ and ${\cal M}_{F,N}$ for the example of $^{76}$Ge.

\item[Fig.5a] Matrix elements ${\cal M}_{GT,N}$ (full line) and
${\cal M}_{F,N}$
(dashed line) for $^{130}$Te as a function of the momentum cutoff
factor $m_A$ [MeV].

\item[Fig.5b] As fig. 5.a, but for
${\cal M}_{GT'}$ (full line) and ${\cal M}_{F'}$ (dashed line) and
${\cal M}_{T'}$ (dash-dotted line).

\item[Fig.6a]  Matrix elements ${\cal M}_{GT,N}$ (full line) and
${\cal M}_{F,N}$ (dashed line) for $^{150}$Nd as a function of
the correlation length $l_C$ [fm].

\item[Fig.6b] As fig. 6.a, but for
${\cal M}_{GT'}$ (full line) and ${\cal M}_{F'}$ (dashed line) and
${\cal M}_{T'}$ (dash-dotted line).

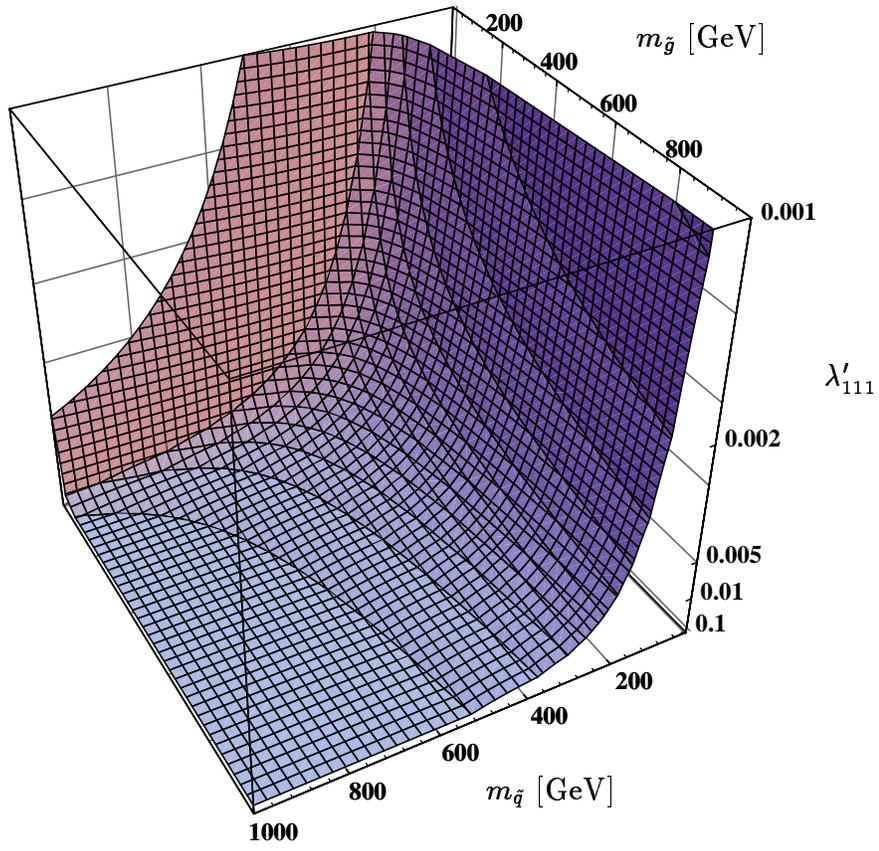
\item[Fig.7] Constraints from $\znbb$ decay on the squark $m_{\tilde q}$ and
gluino $m_{\tilde g}$  masses and
the R-parity violating Yukawa coupling constant $\lambda'_{111}$.
Values on the backside of the surface are forbidden by non-observation of
$\znbb$ decay.
$m_{{\tilde d}_R} = m_{{\tilde u}_L} = m_{\tilde q}$ is assumed
and the matrix element ${\cal M}_{\tilde q}$
for Set A of the coefficients $\alpha^{(i)}$ has been used.

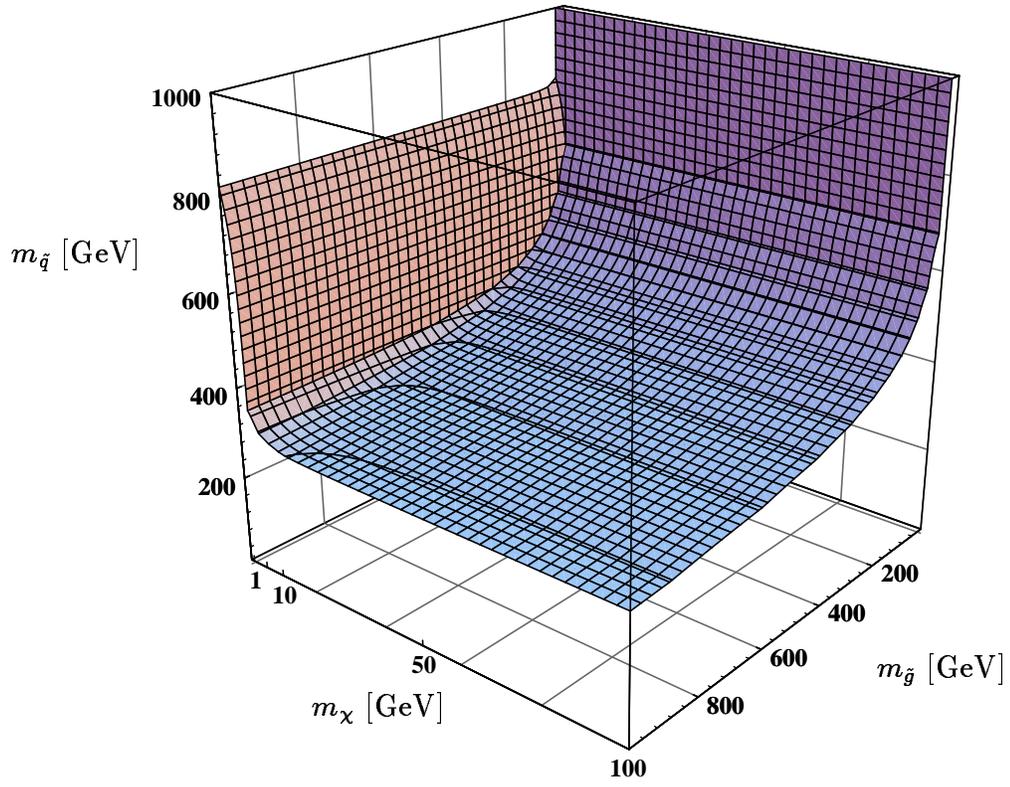
\item[Fig.8] Dependence of the squark and gluino mass limits
derived from $\znbb$ decay on the
actual value of the neutralino mass $m_{\chi}$ for $\lambda'_{111} = 10^{-2}$.

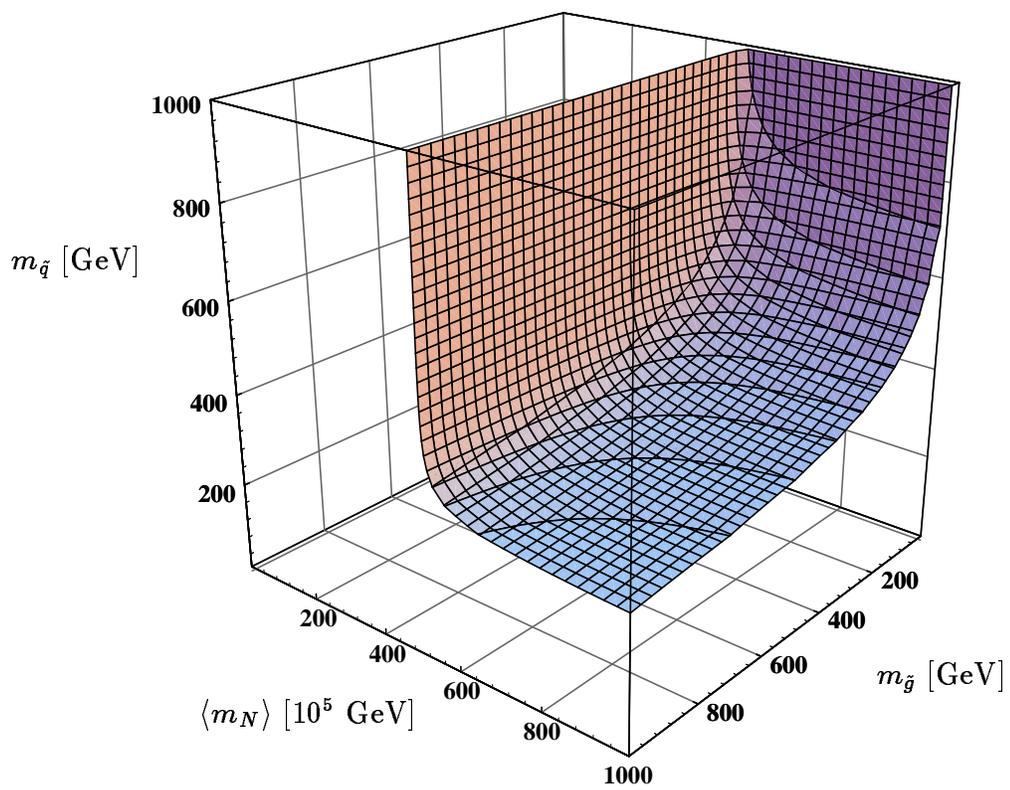
\item[Fig.9] Combined limits on supersymmetric particle masses
$m_{\tilde q, \tilde g}$ and
the effective heavy Majorana neutrino mass $\langle m_N \rangle$
for $\lambda'_{111}$ $=$ $10^{-2}$.

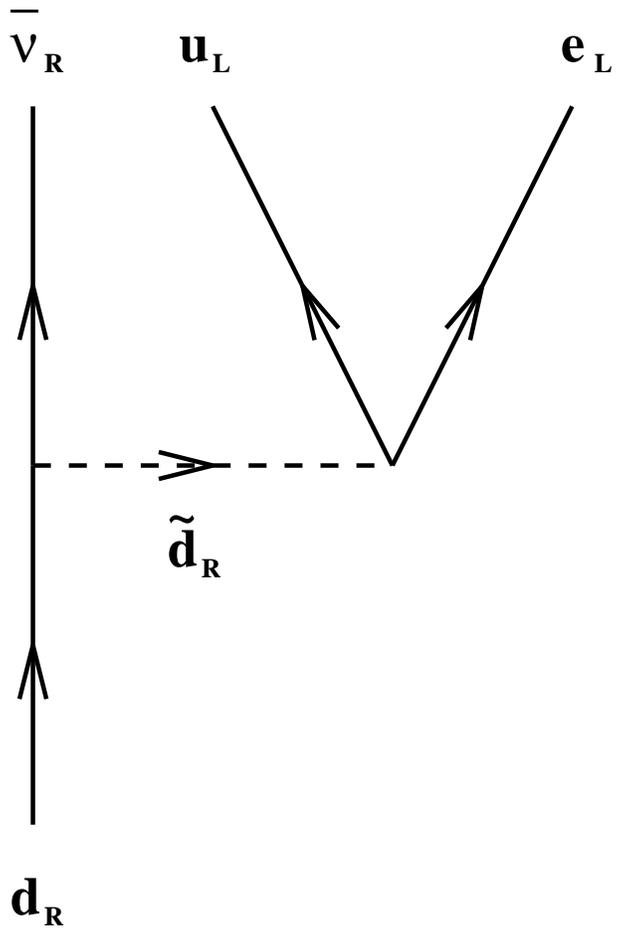
\item[Fig.10]Feynman graph for neutron decay in $R_p$-violating
supersymmetric theories, see text.

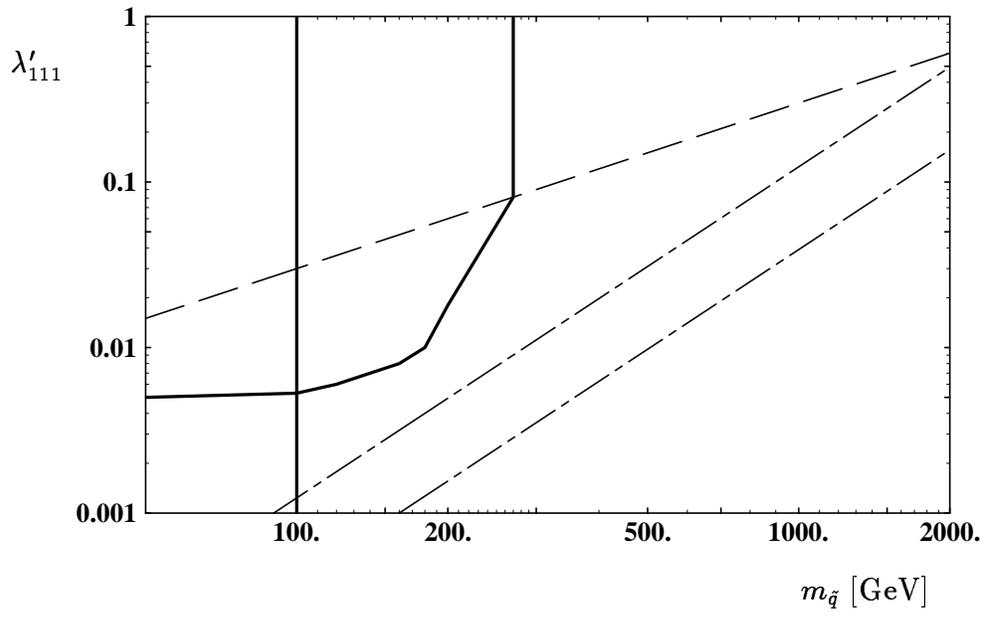
\item[Fig.11] Comparison of limits on R-parity violating supersymmetric
theories from different experiments in the ($m_{\tilde q}$-$\lambda'_{111}$)
plane. The dashed line is the limit from
charged-current universality according to \cite{bar89}. The vertical
line is the lower limit on squark masses in $R_p$-violating
supersymmetric theories from the TEVATRON, according to \cite{roy92}.
The thick full
line is the region which might be explored by HERA with about one year
of data \cite{but93}. The 2 full lines to the right are the limits
obtained from non-observation of the $\znbb$ decay of $^{76}$Ge for
gluino masses of (from left to right) $m_{\tilde g}=$ $1$ TeV,
$100$ GeV, respectively. The parameter regions to the
upper left of the lines are forbidden by the different experiments.
It is seen that even for a gluino mass as large as $1$ TeV, $\znbb$
decay gives the most stringent limits.

\end{itemize}

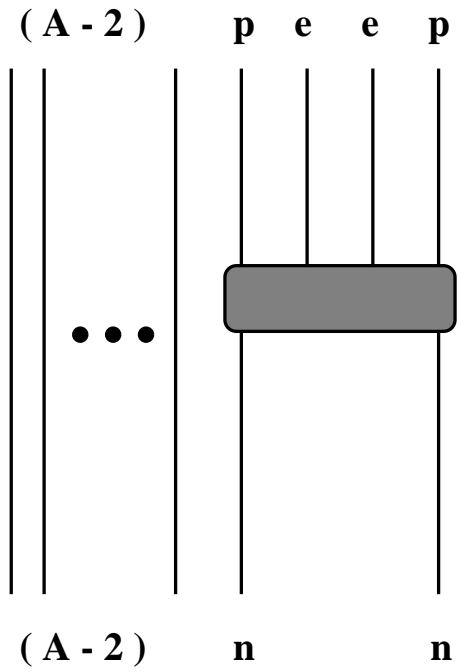
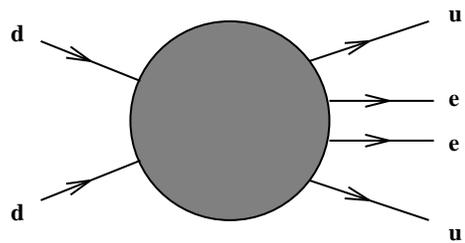
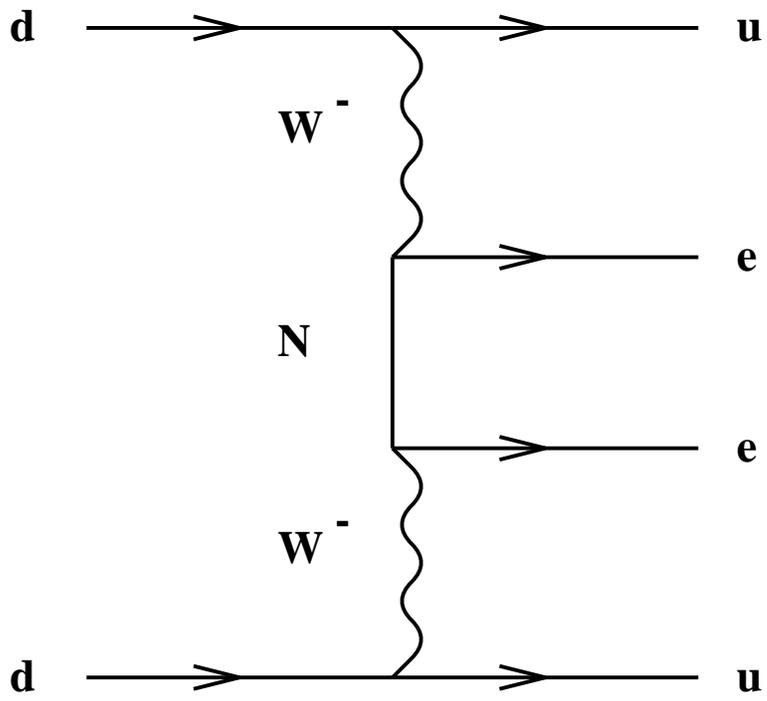
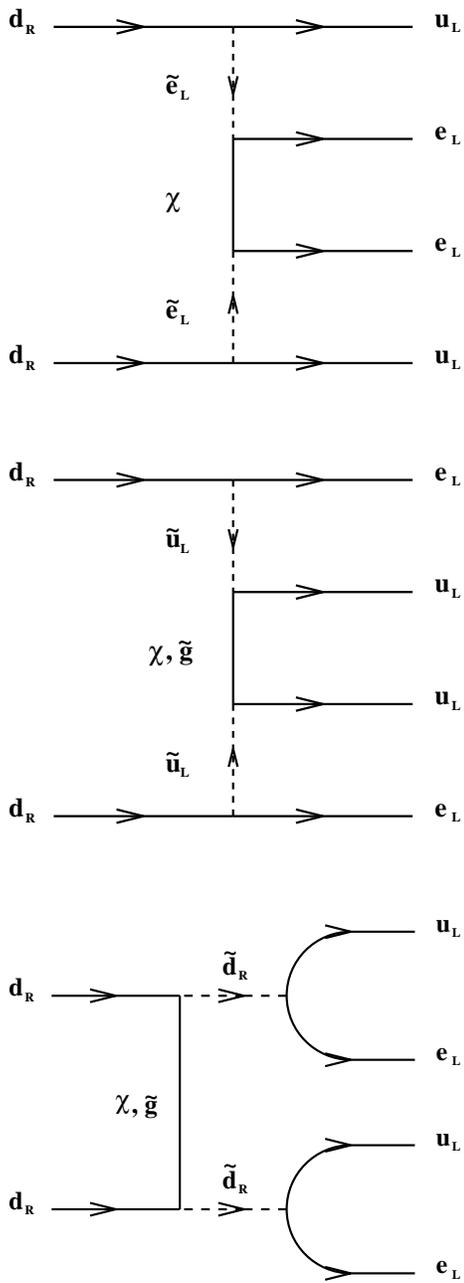
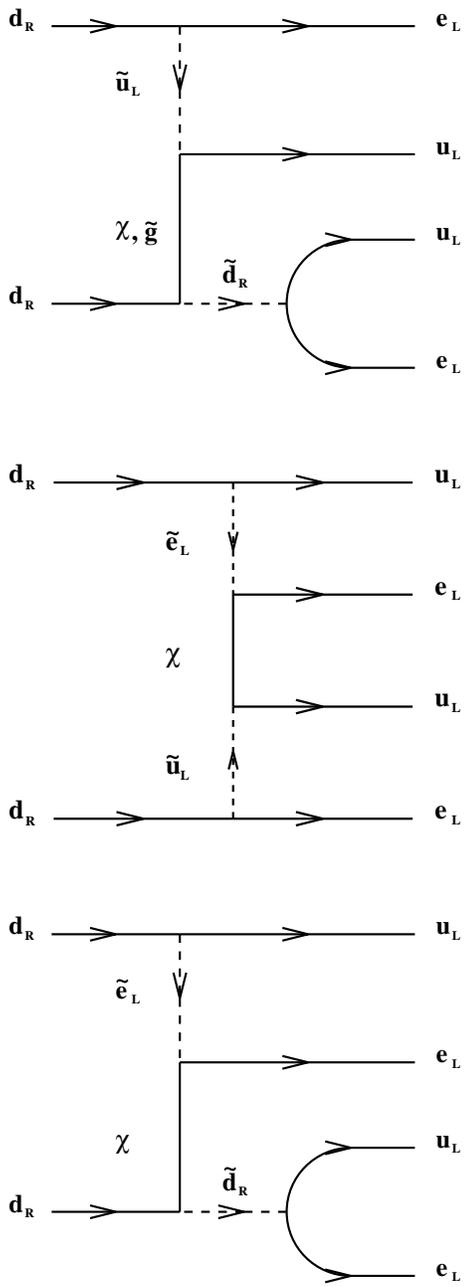
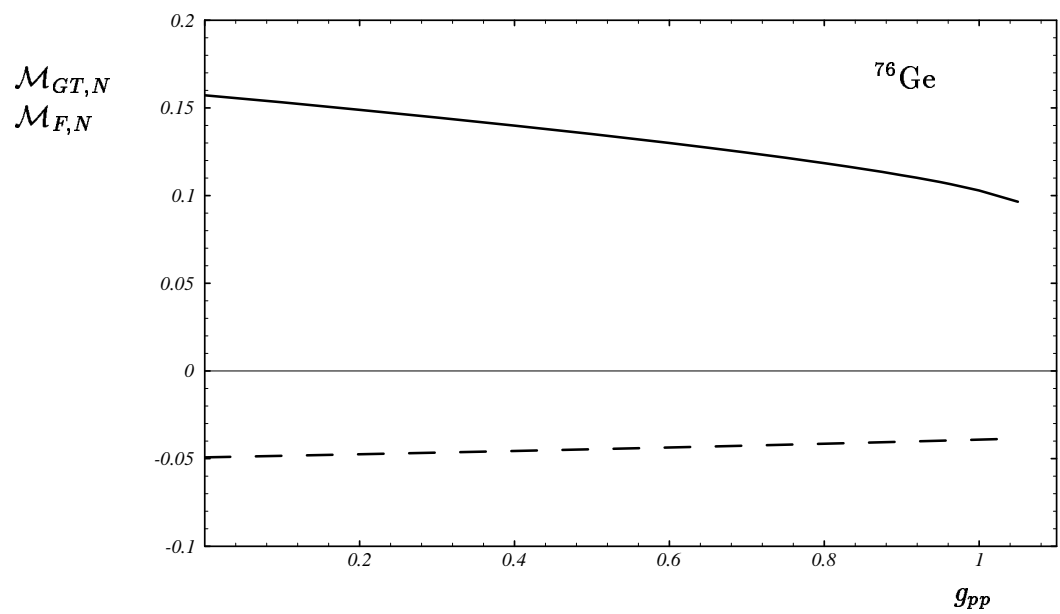
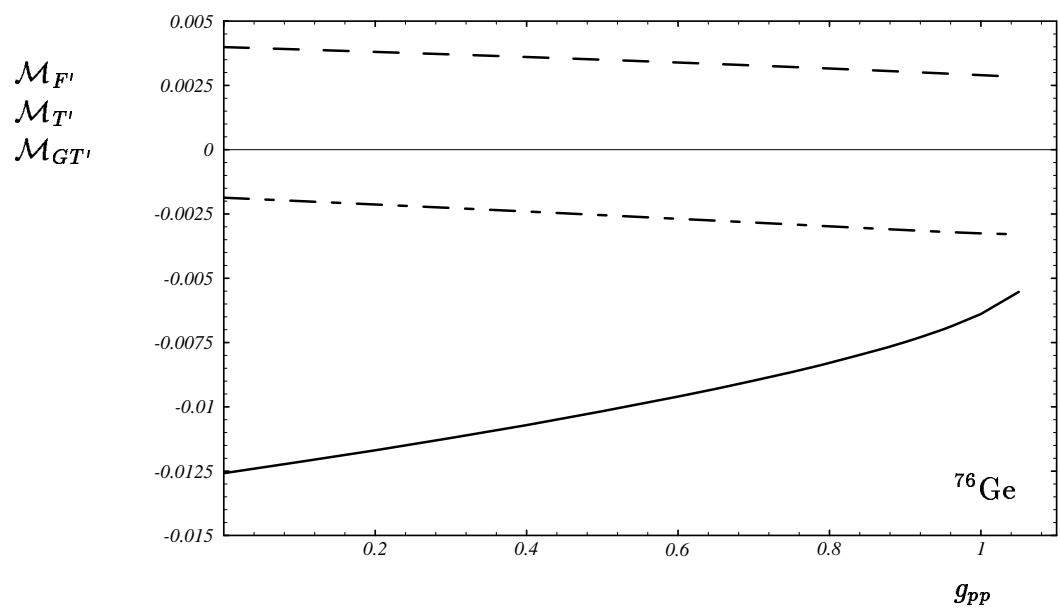
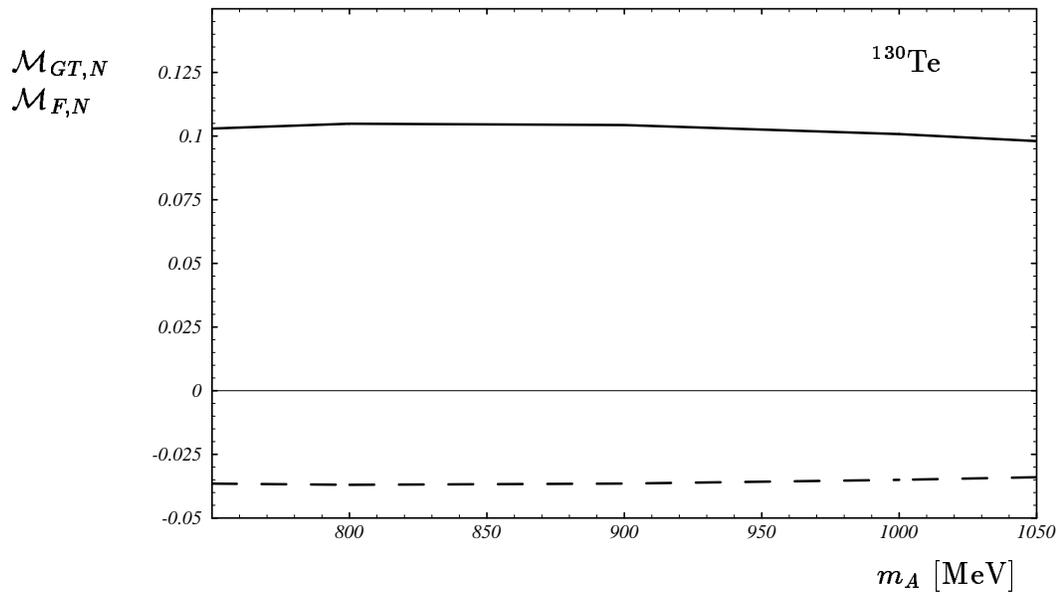
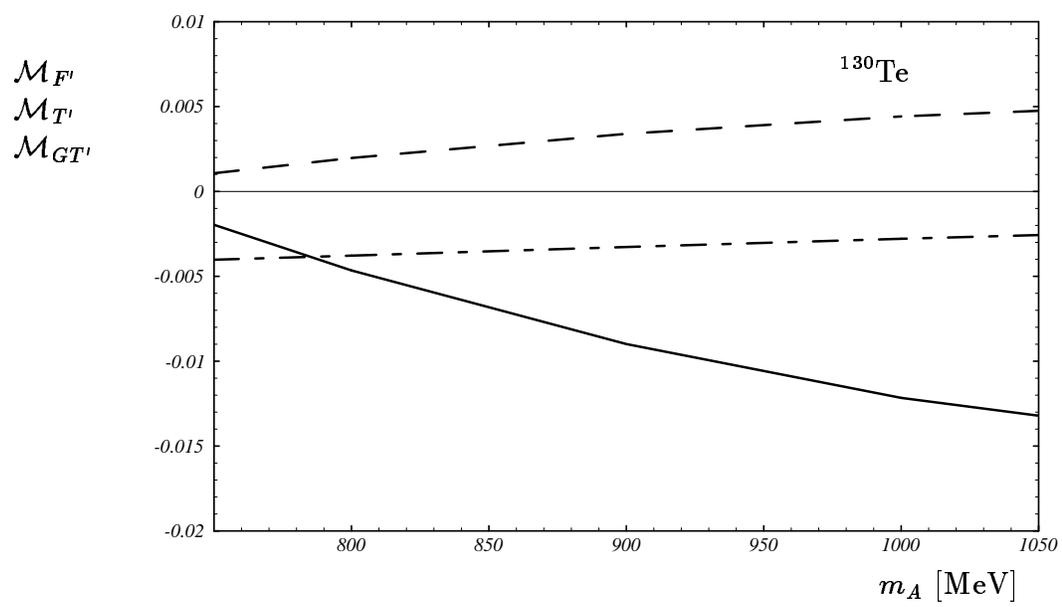
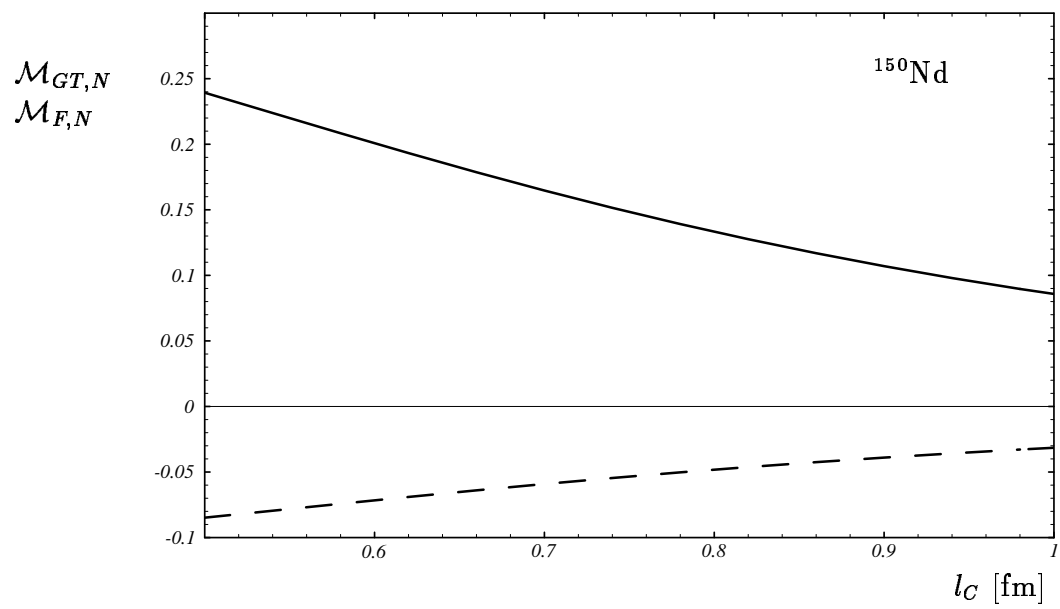
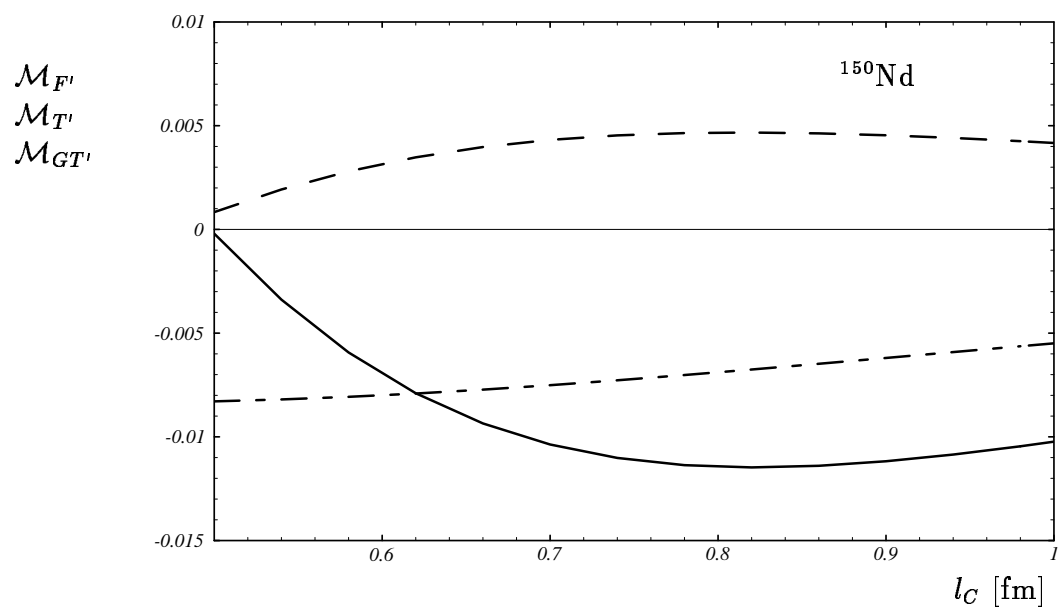
\end{document}